\documentclass[preprint,12pt,a4paper]{elsarticle}

\makeatletter
\def\ps@pprintTitle{%
  \let\@oddhead\@empty
  \let\@evenhead\@empty
  \def\@oddfoot{\reset@font\hfil\thepage\hfil}
  \let\@evenfoot\@oddfoot
}
\makeatother

\usepackage{amssymb}
\usepackage{hyperref}
\usepackage{breakurl}

\usepackage[fleqn]{amsmath}
\begin{document}

\begin{frontmatter}




\title{Scaling of global properties of fluctuating streamwise velocities in pipe flow: Impact of the viscous term}


\author[au1]{Nils T. Basse}
\ead{nils.basse@npb.dk}

\address[au1]{Independent Scientist \\ Trubadurens v\"ag 8, 423 41 Torslanda, Sweden \\ \vspace{10 mm} \small {\rm \today}}

%
%\begin{keyword}
%
%Flow \sep turbulence \sep fluid mechanics \sep plasmas \sep gases \sep liquids \sep mixtures
%
%%% keywords here, in the form: keyword \sep keyword
%
%%% PACS codes here, in the form: \PACS code \sep code
%
%%% MSC codes here, in the form: \MSC code \sep code
%%% or \MSC[2008] code \sep code (2000 is the default)
%
%\end{keyword}

\begin{abstract}
We extend the procedure outlined in [Basse, "Scaling of global properties of fluctuating and mean streamwise velocities in pipe flow: Characterization of a high Reynolds number transition region," Phys. Fluids {\bf 33}, 065127 (2021)] to study global, i.e. radially averaged, scaling of streamwise velocity fluctuations. A viscous term is added to the log-law scaling which leads to the existence of a mathematical abstraction which we call the "global peak". The position and amplitude of this global peak are characterized and compared to the inner and outer peaks. A transition at a friction Reynolds number of order 10000 is identified. Consequences for the global peak scaling, length scales, non-zero asymptotic viscosity, turbulent energy production/dissipation and turbulence intensity scaling are appraised along with the impact of including an additional wake term.
\end{abstract}

\end{frontmatter}



\section{Introduction}

Global, i.e. radially averaged, log- and power-law models for the mean and fluctuating parts of streamwise velocities in pipe flow have been presented in \cite{basse_a} based on Princeton Superpipe measurements \cite{hultmark_a}. In our paper, we treated log- and power-law models with two fit parameters and explained how this could be extended to e.g. three fit parameters. Here, we provide a first example of this using the log-law for streamwise velocity fluctuations, but with an additional viscous term as introduced in \cite{perry_a,perry_b}. The standard expression for streamwise velocity fluctuations is the two-parameter log-law which is a consequence of the attached eddy hypothesis \cite{townsend_a,smits_a,marusic_a}.

One important motivation for the study is that we saw indications of a possible Reynolds number dependence of the streamwise fluctuating velocity, see e.g. Figure 4 in \cite{basse_a}.

The paper is organized as follows: In Section \ref{sec:local} we review local scaling of streamwise velocity fluctuations with an additional viscous term. The radial averaging definitions which are applied to fluctuating velocities are presented in Section \ref{sec:averaging}. The global scaling results are contained in Section \ref{sec:global}. We discuss our findings in Section \ref{sec:discussion} and conclude in Section \ref{sec:conclusions}.

\section{Local scaling}
\label{sec:local}

We use an equation for the square of the normalised fluctuating velocity $u$ including the viscous term $V$ as formulated in \cite{perry_b}:

\begin{eqnarray}
\label{eq:u_rms_sq}
\label{eq:fluc_sq_perry}
 \frac{{\overline{u^2_l}}(z)}{U_{\tau}^2} &=& B_{l} - A_{l} \log (z/\delta) - C_l (z^+)^{-1/2} \\
   &=& B_{l} - A_{l} \log (z/\delta) + V(z^+),
\end{eqnarray}

\noindent where overbar is time averaging, $U_{\tau}$ is the friction velocity, $z$ is the distance from the wall, $\delta$ is the boundary layer thickness (pipe radius $R$ for pipe flow) and $z^+=zU_{\tau}/\nu$ is the normalized distance from the wall, where $\nu$ is the kinematic viscosity. Note that:

\begin{equation}
%\label{}
z/\delta = \frac{z^+}{Re_{\tau}},
\end{equation}

\noindent where $Re_{\tau}=\delta U_{\tau}/\nu$ is the friction Reynolds number.

\begin{figure}[!ht]
\centering
\includegraphics[width=6.5cm]{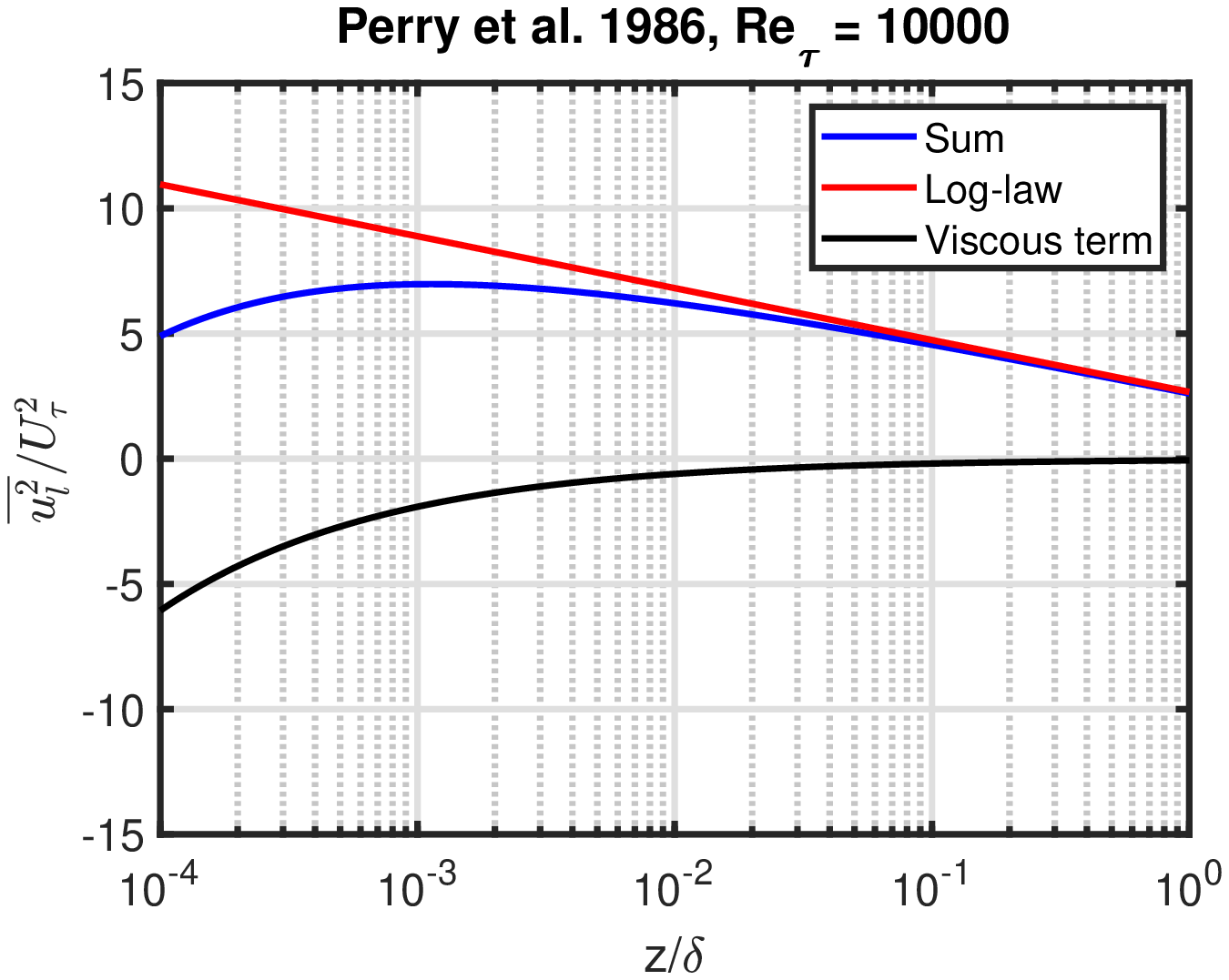}
\hspace{0.3cm}
\includegraphics[width=6.5cm]{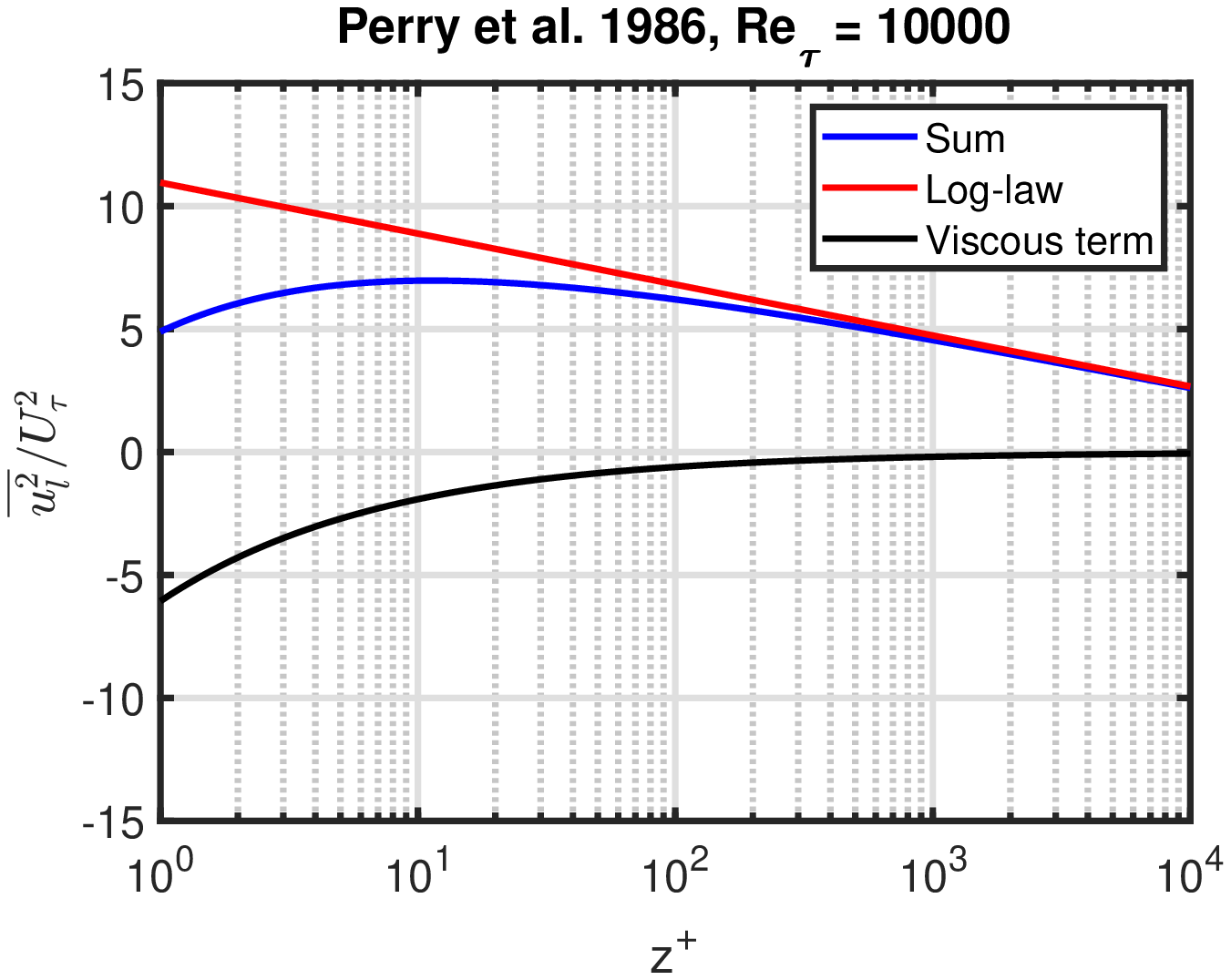}
\caption{$\frac{{\overline{u^2_l}}(z)}{U_{\tau}^2}$ for $Re_{\tau}=10000$ using fit parameters from \cite{perry_b} as a function of $z/\delta$ (left-hand plot) and $z^+$ (right-hand plot).}
\label{fig:fluc_perry_prof}
\end{figure}

In \cite{perry_b}, the fit parameters $A_{l}=0.90$, $B_{l}=2.67$, $C_l=6.06$, see Figure \ref{fig:fluc_perry_prof} for an example. Here, $A_l$ and $C_l$ are universal constants and $B_l$ is a "large-scale characteristic constant", i.e. dependent on the specific geometry. The subscript "l" means that the constants are "local" fits, i.e. a range of $z$ where the model is valid ($z^+>50$ \cite{perry_c}).

Similar \cite{spalart_a} and more complex \cite{marusic_b} viscous terms $V(z^+)$ have been treated previously. We use a simple term for clarity of exposition and to show the qualitative behaviour.

The viscous term has an exponent of -1/2, which is close to what we found when fitting a two-parameter power-law to the radially averaged measurements \cite{basse_a}:

\begin{equation}
\label{eq:pow_m0p48}
\frac{{\overline{u^2_g}}(z)}{U_{\tau}^2} \bigg \rvert_{\rm power-law} = [1.24 \pm 0.04] \times \left( \frac{z}{\delta} \right)^{-0.48 \pm 0.01},
\end{equation}

\noindent where the subscript "g" means that the parameters are "global", i.e. covering the entire range of $z$. We will return to this exponent when we discuss results of including both the viscous and wake \cite{marusic_b,marusic_c,kunkel_a} terms in \ref{sec:app_wake}.

Adding a viscous term means that what we term a "global peak" will exist at a normalized distance from the wall given by:

\begin{eqnarray}
%\nonumber % Remove numbering (before each equation)
\label{eq:z_plus_global_peak}
  z^+ \rvert_{\rm global~peak} &=& \left( \frac{C_l}{2A_{l}} \right)^2 \\
   &=& 11 \label{eq:eleven}
\end{eqnarray}

This value is closer to the wall than the expected validity of the model ($z^+>50$), so we interpret this peak as a mathematical abstraction in some sense. However, it is interesting to note that the position is quite close to what has been measured for the so-called "inner peak" \cite{samie_a}:

\begin{equation}
\label{eq:fifteen}
z^+ \rvert_{\rm inner~peak} = 15
\end{equation}

\begin{figure}[!ht]
\centering
\includegraphics[width=12cm]{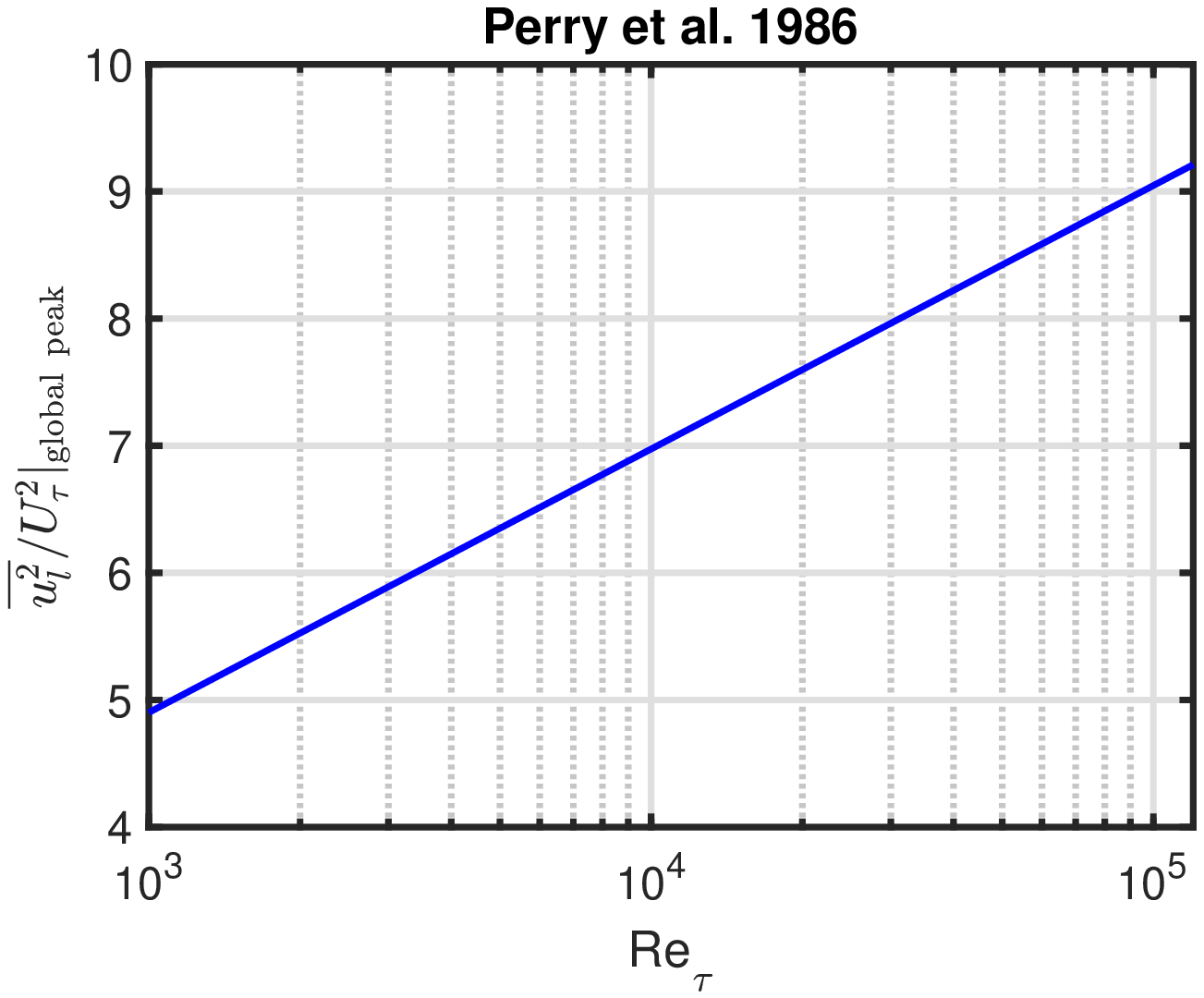}
\caption{$\frac{{\overline{u^2_l}}}{U_{\tau}^2} \bigg \rvert_{\rm global~peak}$ as a function of $Re_{\tau}$ using fit parameters from \cite{perry_b}.}
\label{fig:peak_perry}
\end{figure}

Combining Equations (\ref{eq:fluc_sq_perry}) and (\ref{eq:z_plus_global_peak}) we find that the amplitude of the square of the normalised fluctuating velocity scales as:

\begin{eqnarray}
  \frac{{\overline{u^2_l}}}{U_{\tau}^2} \bigg \rvert_{\rm global~peak} &=& B_{l}-2A_{l} \times (1+\log (C_l)-\log(2A_{l})) + A_{l} \log (Re_{\tau}) \\
   &=&  0.90 \times \log (Re_{\tau}) - 1.32, \label{eq:peak_amp_perry}
\end{eqnarray}

\noindent see Figure \ref{fig:peak_perry}.

\section{Radial averaging definitions}
\label{sec:averaging}

Radial averaging is defined in Equations (\ref{eq:AM_def})-(\ref{eq:VA_end}) for arithmetic mean (AM), area-averaged (AA) and volume-averaged (VA), respectively. The definitions are written both using $z$ and $z^+$.

\begin{eqnarray}
\label{eq:AM_def}
  \langle \cdot \rangle_{\rm AM} &=& \frac{1}{\delta} \int_{0}^{\delta} [\cdot] {\rm d}z \\
   &=& \frac{1}{Re_{\tau}} \int_{0}^{Re_{\tau}} [\cdot] {\rm d}z^+
\end{eqnarray}

\begin{eqnarray}
\label{eq:AA_def}
  \langle \cdot \rangle_{\rm AA} &=& \frac{2}{\delta^2} \int_{0}^{\delta} [\cdot] \times (\delta-z) {\rm d}z \\
   &=& \frac{2}{Re_{\tau}} \int_{0}^{Re_{\tau}} [\cdot] {\rm d}z^+ - \frac{2}{Re_{\tau}^2} \int_{0}^{Re_{\tau}} [\cdot] \times z^+ {\rm d}z^+
\end{eqnarray}

\begin{eqnarray}
\label{eq:VA_def}
  \langle \cdot \rangle_{\rm VA} &=& \frac{3}{\delta^3} \int_{0}^{\delta} [\cdot] \times (\delta-z)^2 {\rm d}z \\
  \label{eq:VA_end}
   &=& \frac{3}{Re_{\tau}} \int_{0}^{Re_{\tau}} [\cdot] {\rm d}z^+ + \frac{3}{Re_{\tau}^3} \int_{0}^{Re_{\tau}} [\cdot] \times (z^+)^2 {\rm d}z^+ \\ \nonumber
   & & - \frac{6}{Re_{\tau}^2} \int_{0}^{Re_{\tau}} [\cdot] \times z^+ {\rm d}z^+
\end{eqnarray}

\section{Global scaling results}
\label{sec:global}

\subsection{Measurements}

\begin{figure}[!ht]
\centering
\includegraphics[width=12cm]{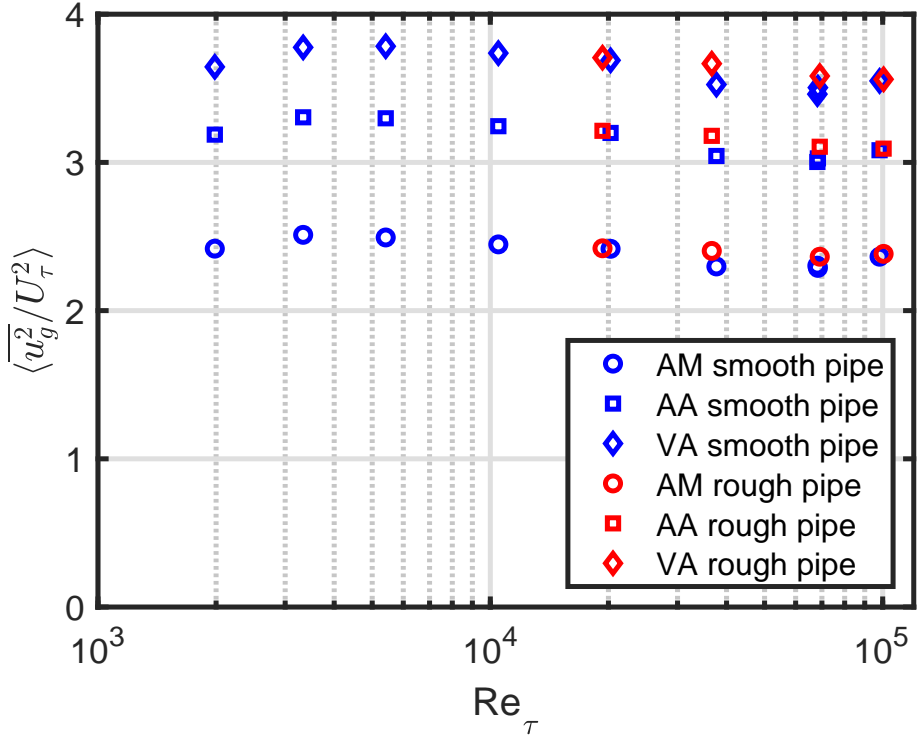}
\caption{The averaged square of the measured normalized fluctuating velocities as a function of Reynolds number.}
\label{fig:fluc_sq_vs_re_tau}
\end{figure}

The Princeton Superpipe measurements we analyze have a maximum $Re_{\tau}=98190$, whereas direct numerical simulations (DNS) currently have maximum values around 6000 \cite{pirozzoli_a,smits_b}. Thus, we continue to rely solely on measurements for $Re_{\tau}>6000$. This is important for our investigation since it will become clear that a transition takes place for $Re_{\tau}>10000$. Globally averaged measurements of the streamwise square of the normalised fluctuating velocity are presented in Figure \ref{fig:fluc_sq_vs_re_tau}. The measurements constitute the foundation of the results presented in the remainder of this paper.

All fits shown in our paper are fits to smooth pipe measurements, the rough pipe measurements are shown for reference.

\subsection{Model}

Analytical integration of Equation (\ref{eq:fluc_sq_perry}) using Equations (\ref{eq:AM_def})-(\ref{eq:VA_end}) yields these three equations with the three unknowns $A_g$, $B_g$ and $C_g$:

\begin{equation}
%\label{}
\biggl \langle \frac{{\overline{u^2_g}}}{U_{\tau}^2} \biggr \rangle_{\rm AM} = B_{g} + A_{g} - \frac{2 C_g}{\sqrt{Re_{\tau}}}
\end{equation}

\begin{equation}
%\label{}
\biggl \langle \frac{{\overline{u^2_g}}}{U_{\tau}^2} \biggr \rangle_{\rm AA} = B_{g} + \frac{3}{2} \times A_{g} - \frac{8 C_g}{3\sqrt{Re_{\tau}}}
\end{equation}

\begin{equation}
%\label{}
\biggl \langle \frac{{\overline{u^2_g}}}{U_{\tau}^2} \biggr \rangle_{\rm VA} = B_{g} + \frac{11}{6} \times A_{g} - \frac{16 C_g}{5\sqrt{Re_{\tau}}}
\end{equation}

The solutions to these three equations are:

\begin{equation}
%\label{}
A_{g}= -12 \times \biggl \langle \frac{{\overline{u^2_g}}}{U_{\tau}^2} \biggr \rangle_{\rm AM} + 27 \times \biggl \langle \frac{{\overline{u^2_g}}}{U_{\tau}^2} \biggr \rangle_{\rm AA} - 15 \times \biggl \langle \frac{{\overline{u^2_g}}}{U_{\tau}^2} \biggr \rangle_{\rm VA}
\end{equation}

\begin{equation}
%\label{}
B_{g}= -2 \times \biggl \langle \frac{{\overline{u^2_g}}}{U_{\tau}^2} \biggr \rangle_{\rm AM} + \frac{21}{2} \times \biggl \langle \frac{{\overline{u^2_g}}}{U_{\tau}^2} \biggr \rangle_{\rm AA} - \frac{15}{2} \times \biggl \langle \frac{{\overline{u^2_g}}}{U_{\tau}^2} \biggr \rangle_{\rm VA}
\end{equation}

\begin{eqnarray}
  C_g &=& \sqrt{Re_{\tau}} \\
   && \times \Bigg( \Bigg. -\frac{15}{2} \times \biggl \langle \frac{{\overline{u^2_g}}}{U_{\tau}^2} \biggr \rangle_{\rm AM} + \frac{75}{4} \times \biggl \langle \frac{{\overline{u^2_g}}}{U_{\tau}^2} \biggr \rangle_{\rm AA} \\
   && - \frac{45}{4} \times \biggl \langle \frac{{\overline{u^2_g}}}{U_{\tau}^2} \biggr \rangle_{\rm VA} \Bigg. \Bigg) \nonumber
\end{eqnarray}

\begin{figure}[!ht]
\centering
\includegraphics[width=6.5cm]{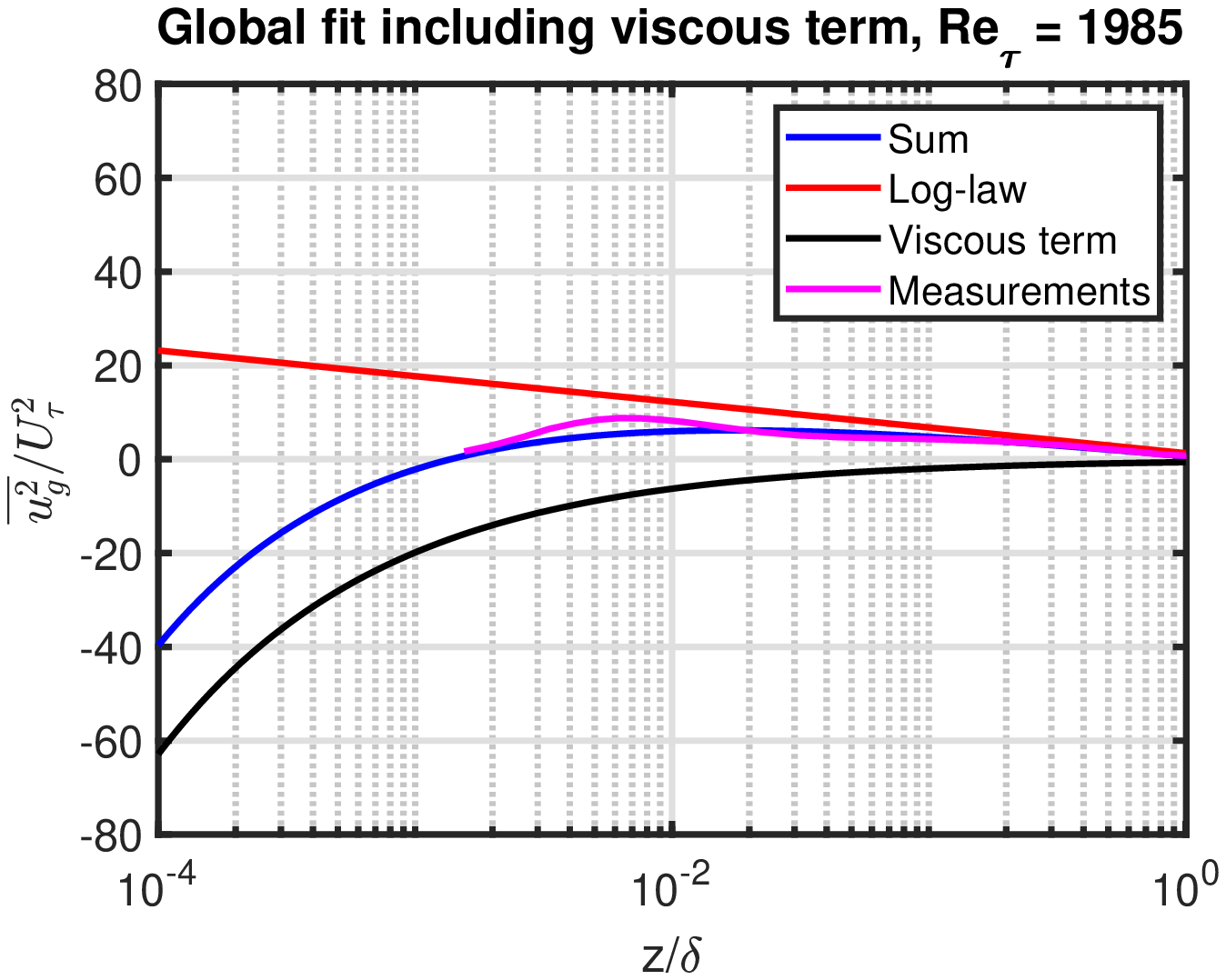}
\hspace{0.3cm}
\includegraphics[width=6.5cm]{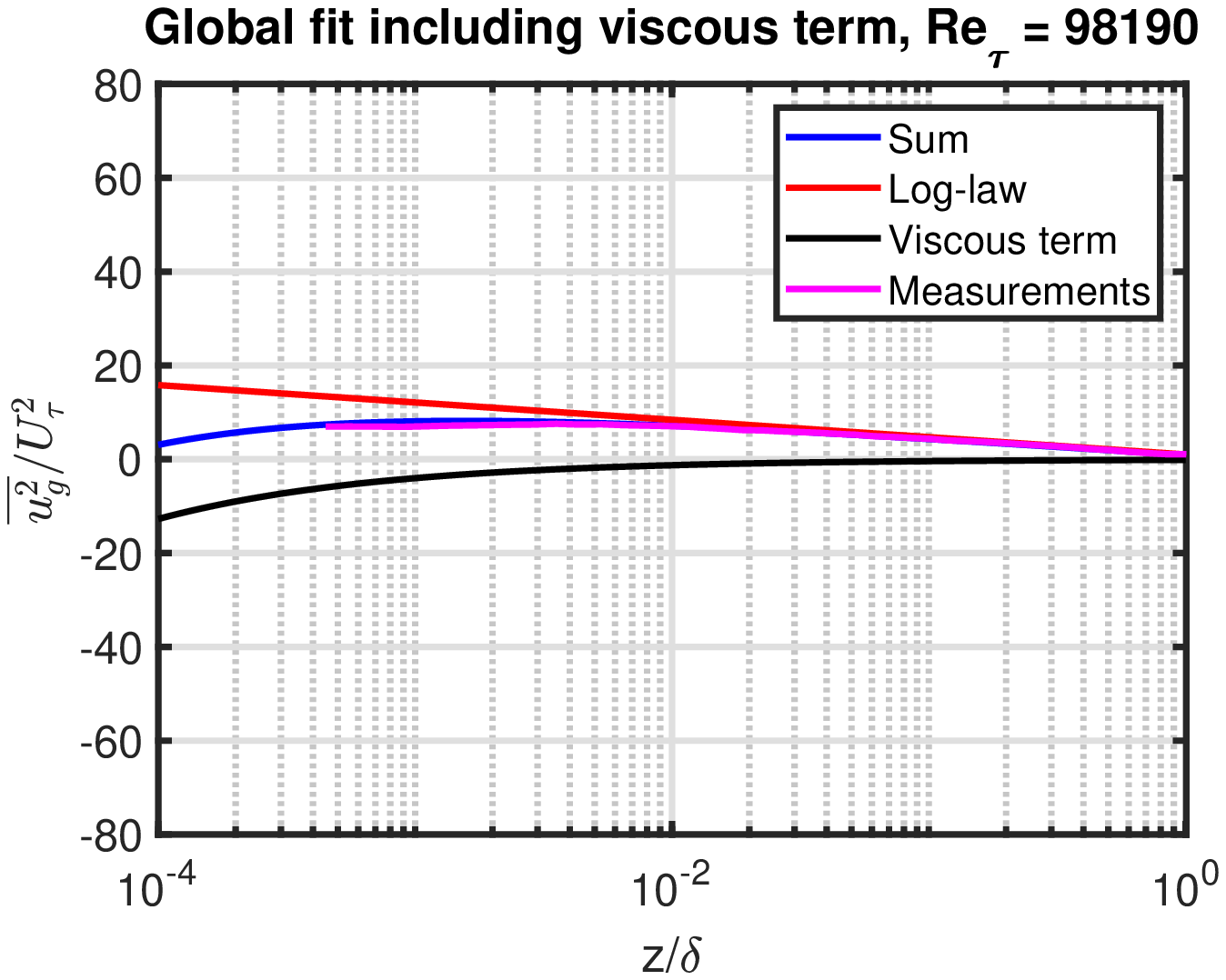}
\caption{$\frac{{\overline{u^2_g}}(z)}{U_{\tau}^2}$ using Equation (\ref{eq:fluc_sq_perry}) and the global fit parameters as a function of $z/\delta$. Left-hand plot: Lowest measured $Re_{\tau}$, right-hand plot: Highest measured $Re_{\tau}$.}
\label{fig:fluc_glob_z_div_delta}
\end{figure}

\begin{figure}[!ht]
\centering
\includegraphics[width=6.5cm]{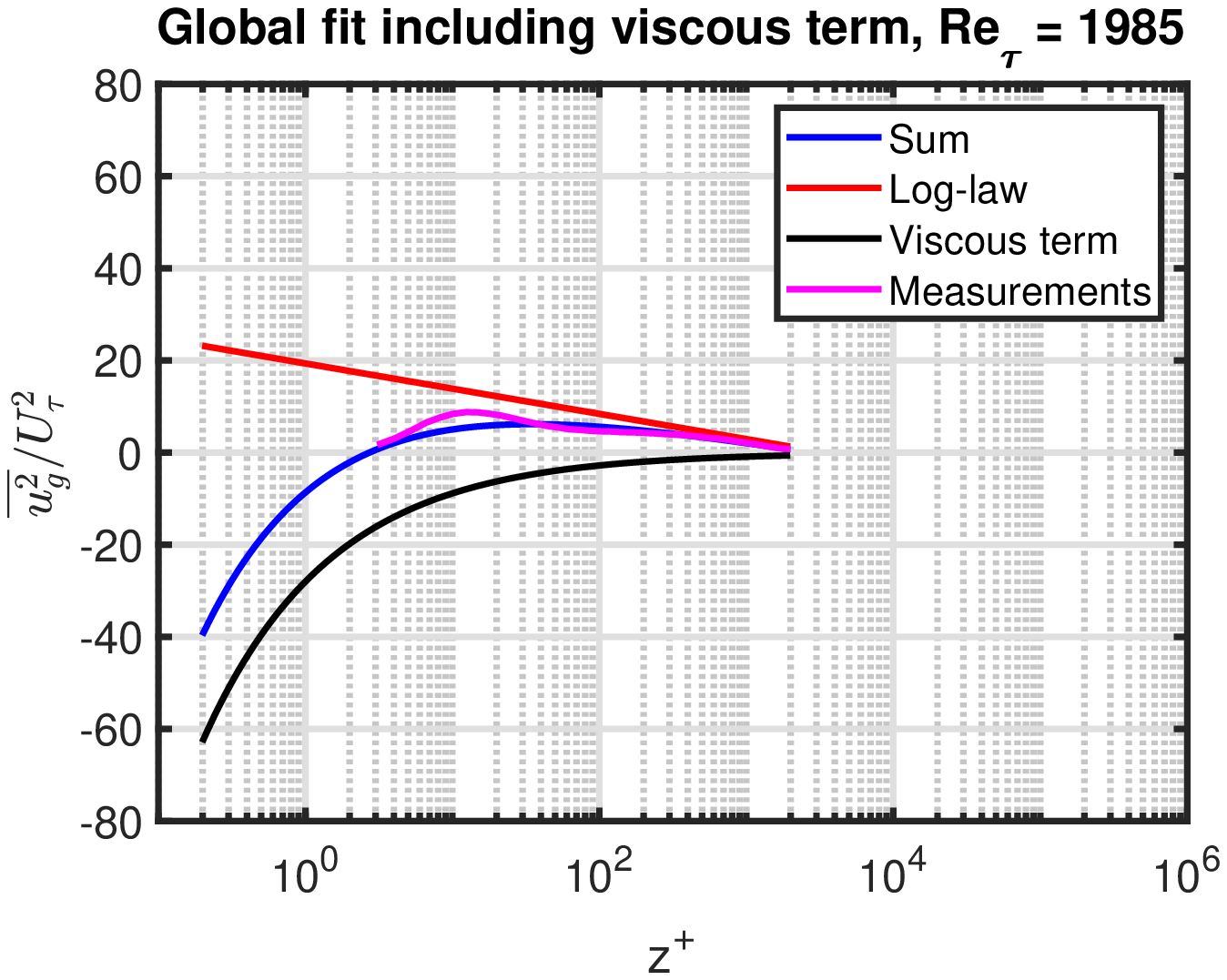}
\hspace{0.3cm}
\includegraphics[width=6.5cm]{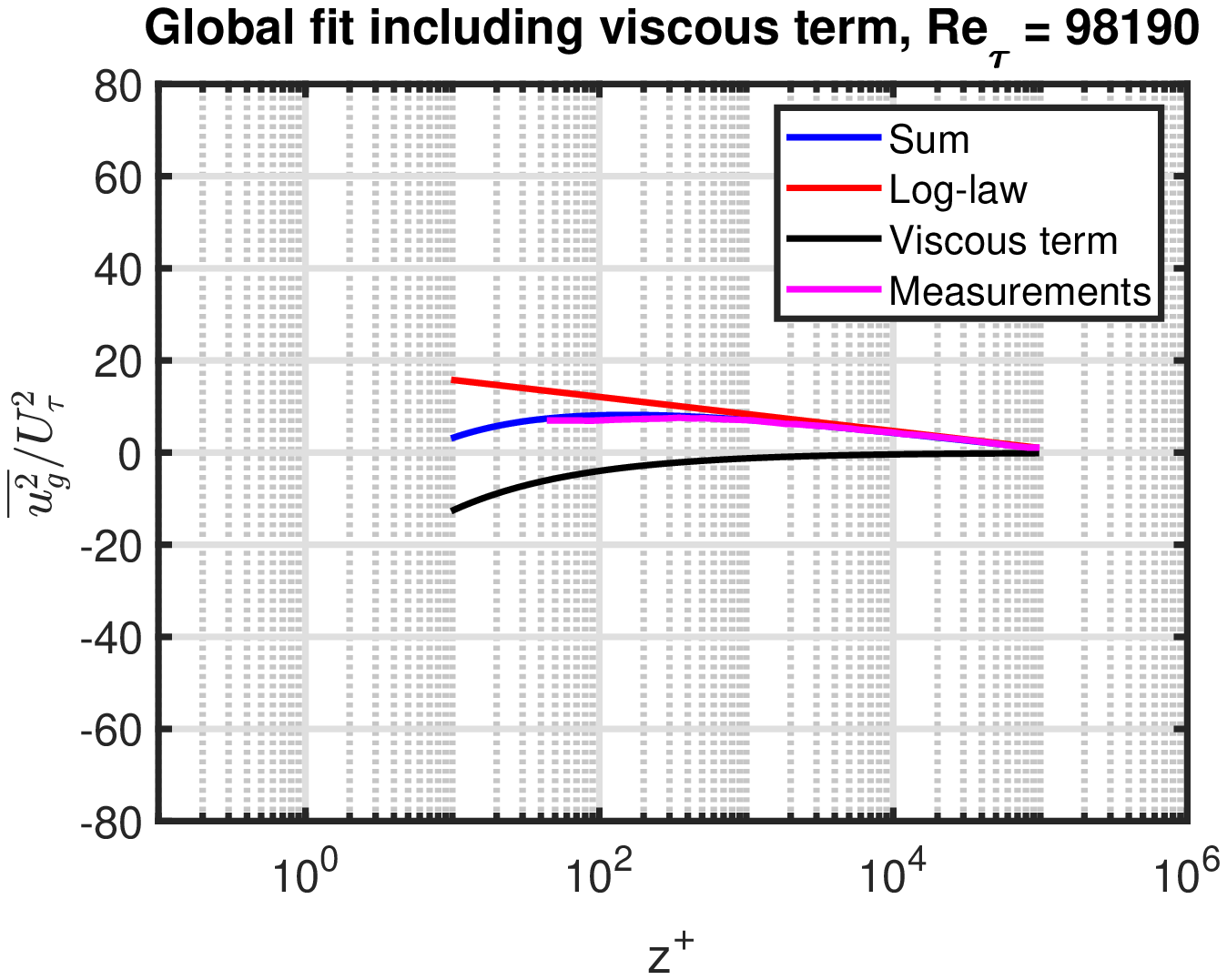}
\caption{$\frac{{\overline{u^2_g}}(z)}{U_{\tau}^2}$ using Equation (\ref{eq:fluc_sq_perry}) and the global fit parameters as a function of $z^+$. Left-hand plot: Lowest measured $Re_{\tau}$, right-hand plot: Highest measured $Re_{\tau}$.}
\label{fig:fluc_glob_z_plus}
\end{figure}

We show examples of the resulting global profiles of the streamwise square of the normalised fluctuating velocity in Figures \ref{fig:fluc_glob_z_div_delta} and \ref{fig:fluc_glob_z_plus} for the lowest ($Re_{\tau}=1985$) and highest ($Re_{\tau}=98190$) Reynolds numbers measured. The change of the log-law part with Reynolds number is modest, but the difference for the viscous term is significant; we will quantify this below. Note that a peak is visible for the low Reynolds number measurements but not visible for the high Reynolds number, probably because the measurements are not available close to the wall. The sum of the log-law and viscous terms is seen to be negative towards the wall for the low Reynolds number case which is the solution to the system of equations in a global sense but not physical in a local sense - locally, our study is a mathematical abstraction.

\subsection{Parameter fits}

\begin{figure}[!ht]
\centering
\includegraphics[width=12cm]{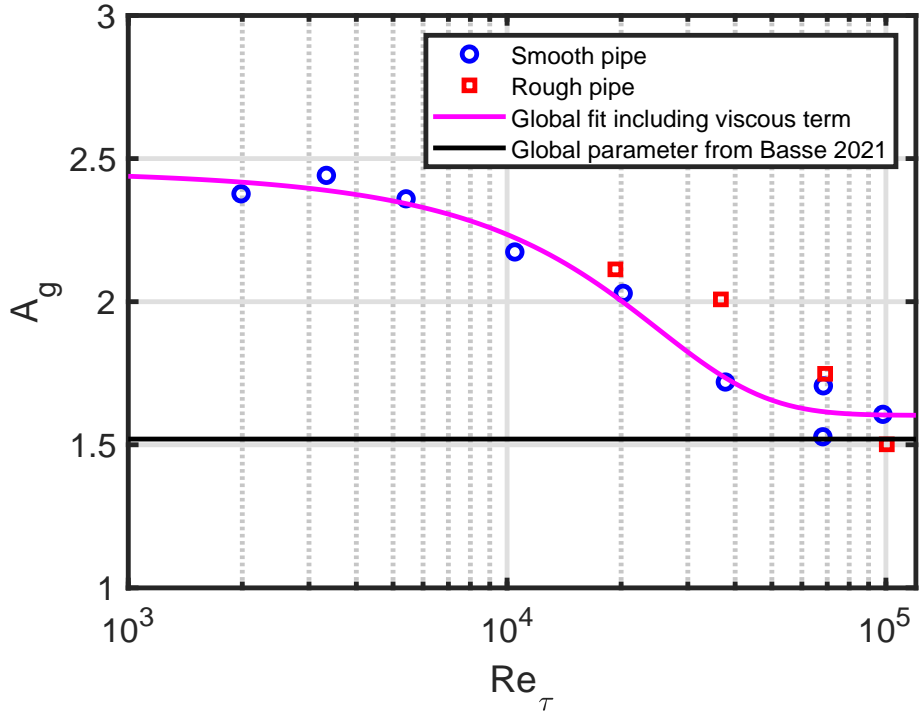}
\caption{$A_g$ as a function of $Re_{\tau}$.}
\label{fig:glob_visc_fit_A_g}
\end{figure}

\begin{figure}[!ht]
\centering
\includegraphics[width=12cm]{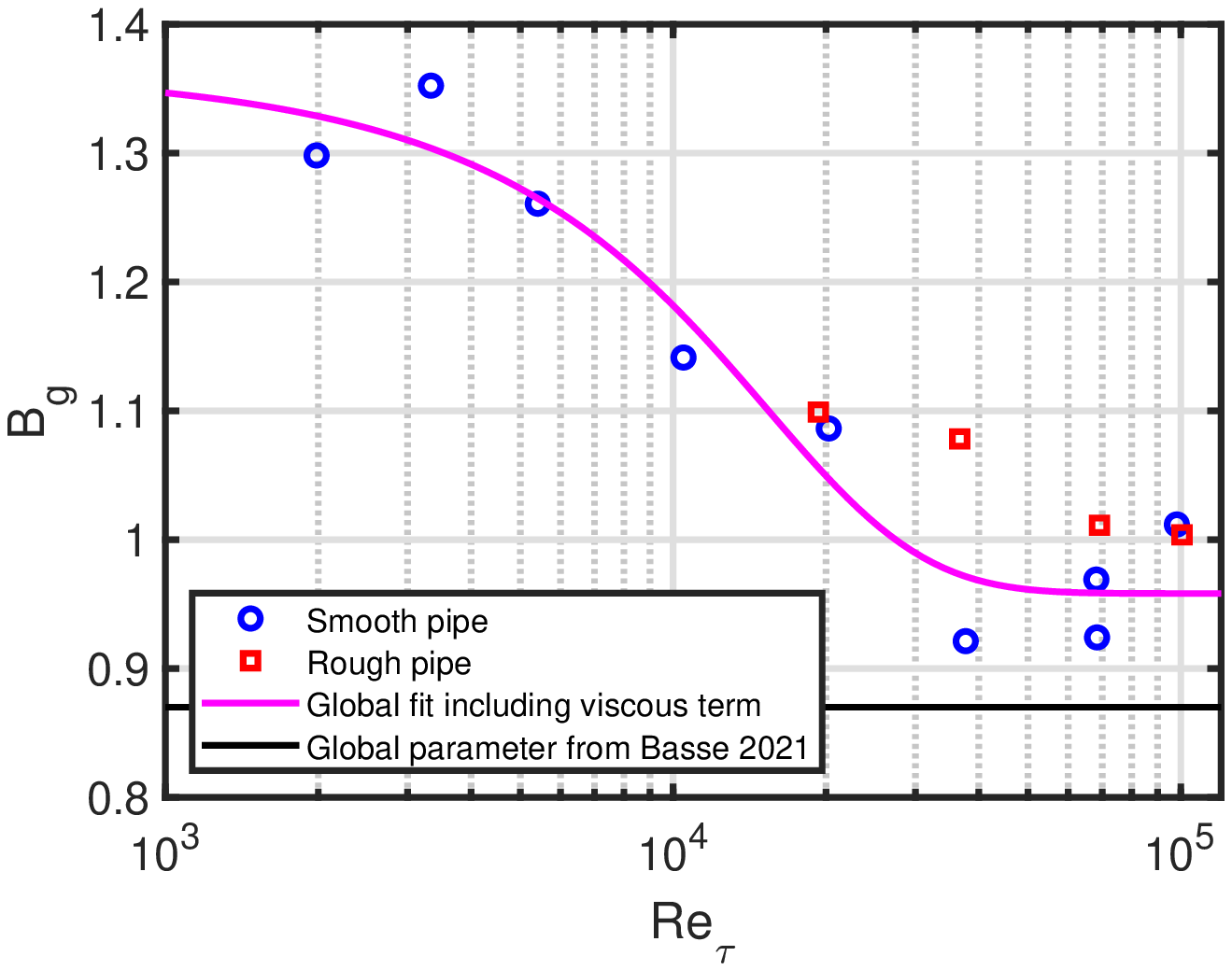}
\caption{$B_g$ as a function of $Re_{\tau}$.}
\label{fig:glob_visc_fit_B_g}
\end{figure}

\begin{figure}[!ht]
\centering
\includegraphics[width=6.5cm]{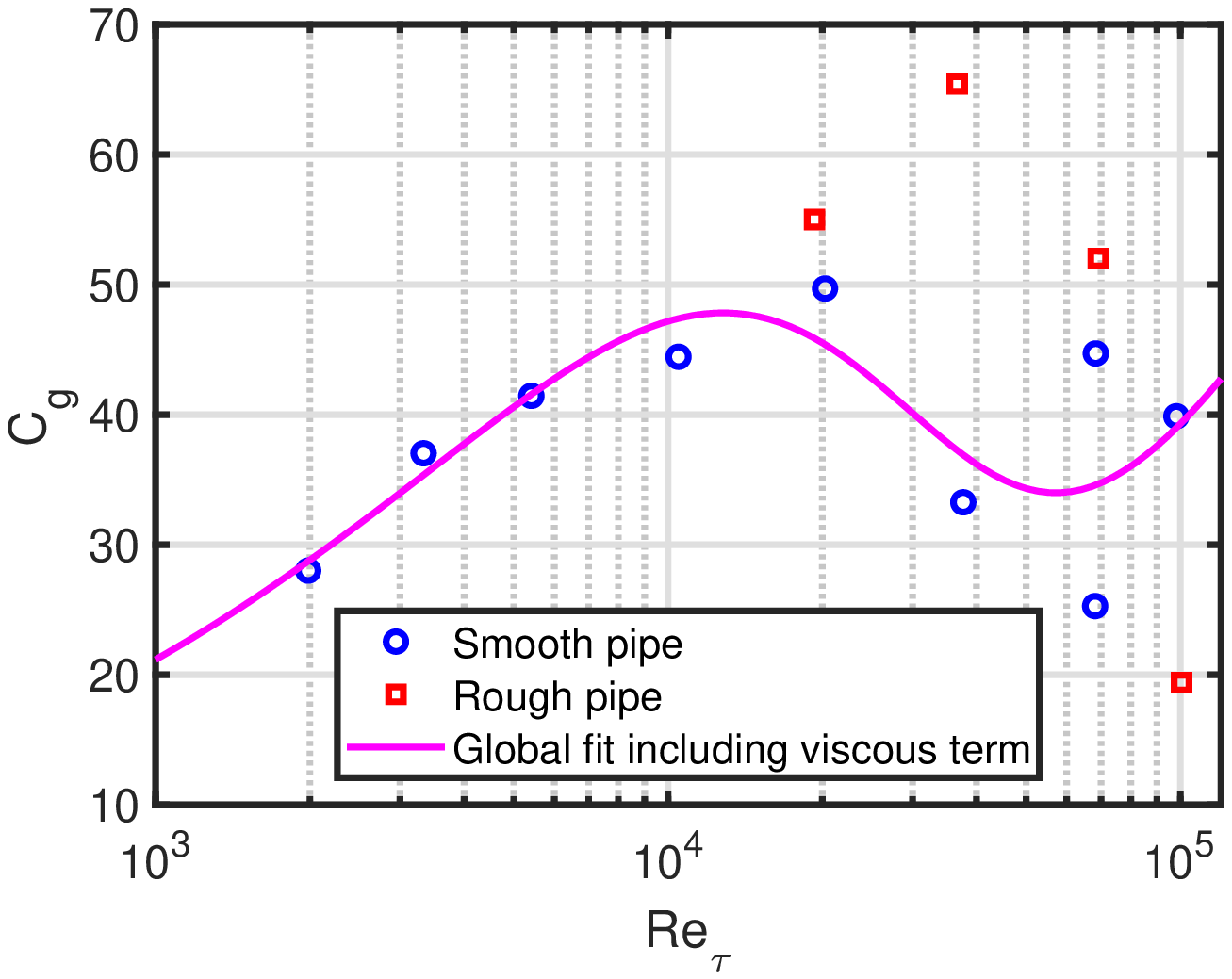}
\hspace{0.3cm}
\includegraphics[width=6.5cm]{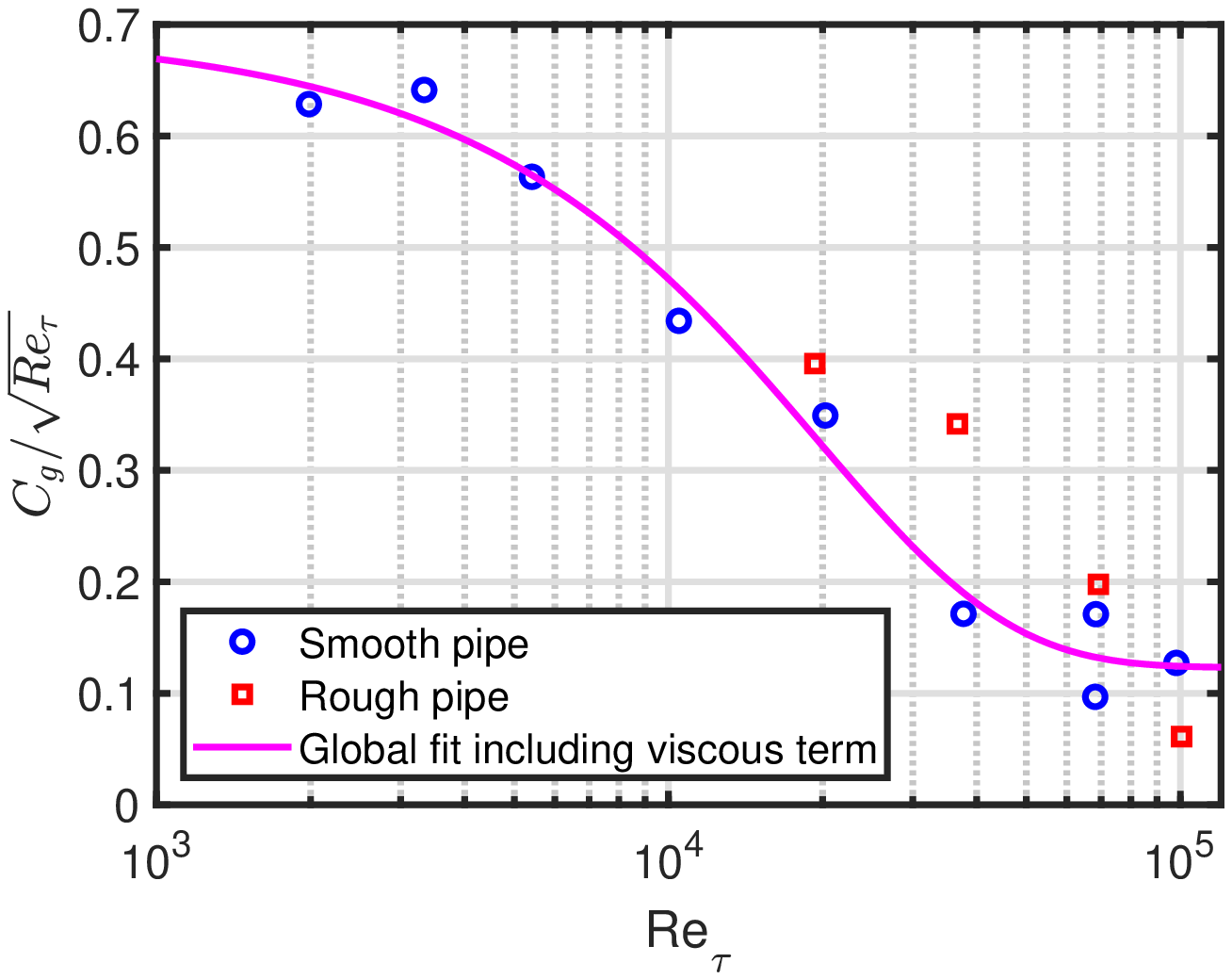}
\caption{Left-hand plot: $C_g$ as a function of $Re_{\tau}$, right-hand plot: $C_g/\sqrt{Re_{\tau}}$ as a function of $Re_{\tau}$.}
\label{fig:glob_visc_fit_C_g}
\end{figure}

We now turn to the topic of whether the global fit parameters scale with $Re_{\tau}$. It is clear from Figures \ref{fig:glob_visc_fit_A_g}-\ref{fig:glob_visc_fit_C_g} that $A_g$, $B_g$ and $C_g/\sqrt{Re_{\tau}}$ scale with $Re_{\tau}$, and that the scaling of $C_g$ is less obvious. In Figures \ref{fig:glob_visc_fit_A_g} and \ref{fig:glob_visc_fit_B_g} we also include the global log-law parameters found in \cite{basse_a}, i.e. $A_{g,{\rm log-law}}=1.52$ and $B_{g,{\rm log-law}}=0.87$.

It is not shown here, but we have attempted both log- and power-law fits to the parameters without finding a satisfactory match. Instead, the best fit is found using an expression including hyperbolic tangent:

\begin{equation}
\label{eq:Q_fit}
Q(Re_{\tau}) = a+b \times \tanh (c \times \left[ Re_{\tau}-d \right]),
\end{equation}

\noindent where $Q$ is the quantity to fit and ($a, b, c, d$) are fit parameters. Note that we have also tested using the tangent function, but hyperbolic tangent is marginally better. The resulting fit parameters are collected in Table \ref{tab:tanh_fit_parameters}; the values for $d$ indicate the transitional Reynolds number which is around 11000-12000 for $A_g$ and $C_g/\sqrt{Re_{\tau}}$ but lower for $B_g$. Perhaps this difference is because $B_g$ depends on the specific geometry, i.e. a pipe for our case.

The coefficient of determination, $R^2$, is also included in the table, where a value of 1 means that the fit matches the data exactly.

The fits are to the smooth wall pipe measurements, but the rough wall measurements are shown for reference.

Asymptotic values for the fit parameters are:

\begin{eqnarray}
  \lim_{Re_{\tau}\to\infty} A_g &=& 1.60 \\
  \lim_{Re_{\tau}\to\infty} B_g  &=& 0.96 \\
  \lim_{Re_{\tau}\to\infty} C_g/\sqrt{Re_{\tau}} &=& 0.12 \label{eq:C_g_div},
\end{eqnarray}

\noindent where Equation (\ref{eq:C_g_div}) implies that:

\begin{equation}
%\label{}
  \lim_{Re_{\tau}\to\infty} C_g = 0.12 \times \sqrt{Re_{\tau}}
\end{equation}

For $C_g$ (left-hand plot in Figure \ref{fig:glob_visc_fit_C_g}) it is interesting to note the non-monotonic variation with $Re_{\tau}$; this appears to be a transition between two different scalings for low and high Reynolds numbers, see Section \ref{subsec:peak_scal}. Possibly the inner peak dominates for lower Reynolds numbers and the outer peak for higher Reynolds numbers, see Section \ref{subsec:peak_scal}.

\hspace{0.5cm}

\begin{table}[!ht]
\caption{Fits to parameters using Equation (\ref{eq:Q_fit}).} % title of Table
\centering % used for centering table
\begin{tabular}{cccccc} % centered columns (4 columns)
\hline\hline %inserts double horizontal lines
Parameter & $a$ & $b$ & $c$ & $d$ & $R^2$\\  % inserts table
%heading
\hline % inserts single horizontal line
$A_g$ & 2.21 & -0.60 & 3.97e-5 & 11186 & 0.98\\
$B_g$ & 1.28 & -0.32 & 5.85e-5 & 4609 & 0.94\\
$C_g/\sqrt{Re_{\tau}}$ & 1.03 & -0.91 & 3.30e-5 & -11755 & 0.98\\
\hline %inserts single line
\end{tabular}
\label{tab:tanh_fit_parameters} % is used to refer this table in the text
\end{table}

\subsection{Average fits}

\begin{figure}[!ht]
\centering
\includegraphics[width=6.5cm]{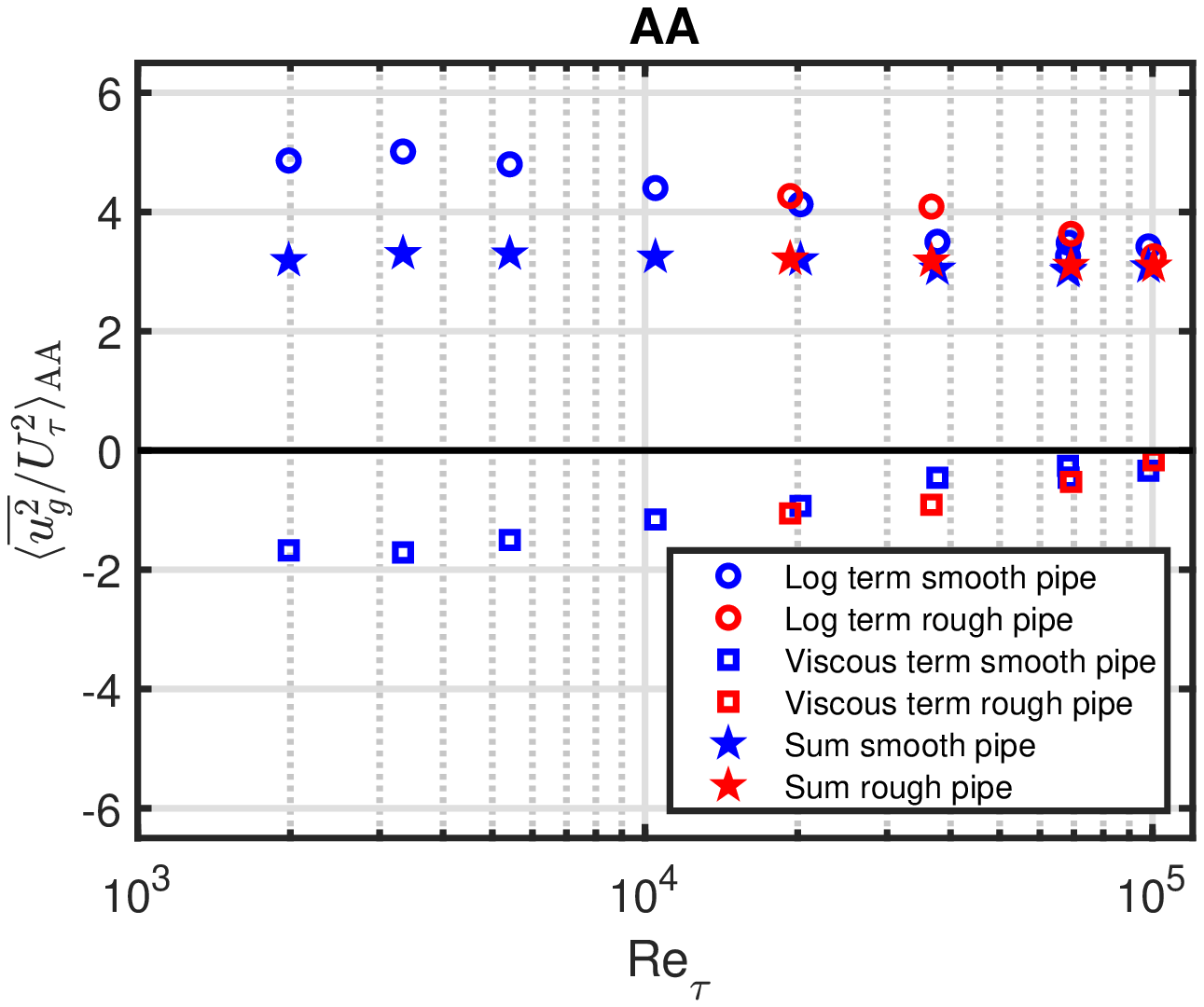}
\hspace{0.3cm}
\includegraphics[width=6.5cm]{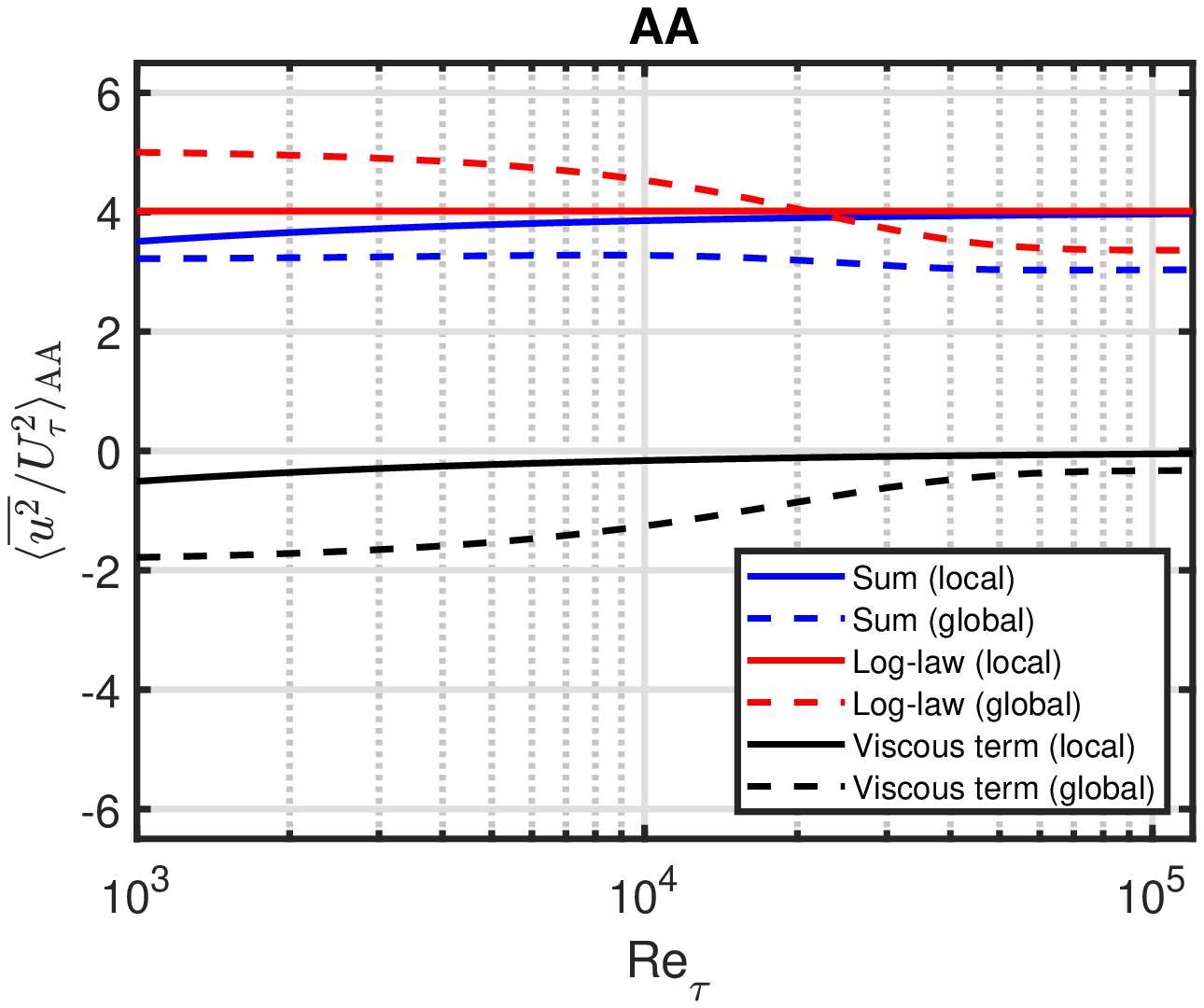}
\caption{Decomposition of the AA average as a function of $Re_{\tau}$. Left-hand plot: Individual points, right-hand plot: Comparison of local and global fits.}
\label{fig:fluc_AA_avrg}
\end{figure}

The decomposition of the fluctuations into a log-law and a viscous term is shown for the AA case in Figure \ref{fig:fluc_AA_avrg}, both for individual points (left-hand plot) and fits (right-hand plot). It is clear that the viscous term decreases with increasing Reynolds number, i.e. the viscosity contribution becomes less important. The magnitude of the viscous term is much larger for the global fit than for the local fit and has a non-zero asymptotic value. The global log-law term decreases with increasing Reynolds number and crosses the local log-law term at $Re_{\tau} \sim 20000$. We conclude that the main difference is due to the behaviour of the viscous term.

\begin{figure}[!ht]
\centering
\includegraphics[width=12cm]{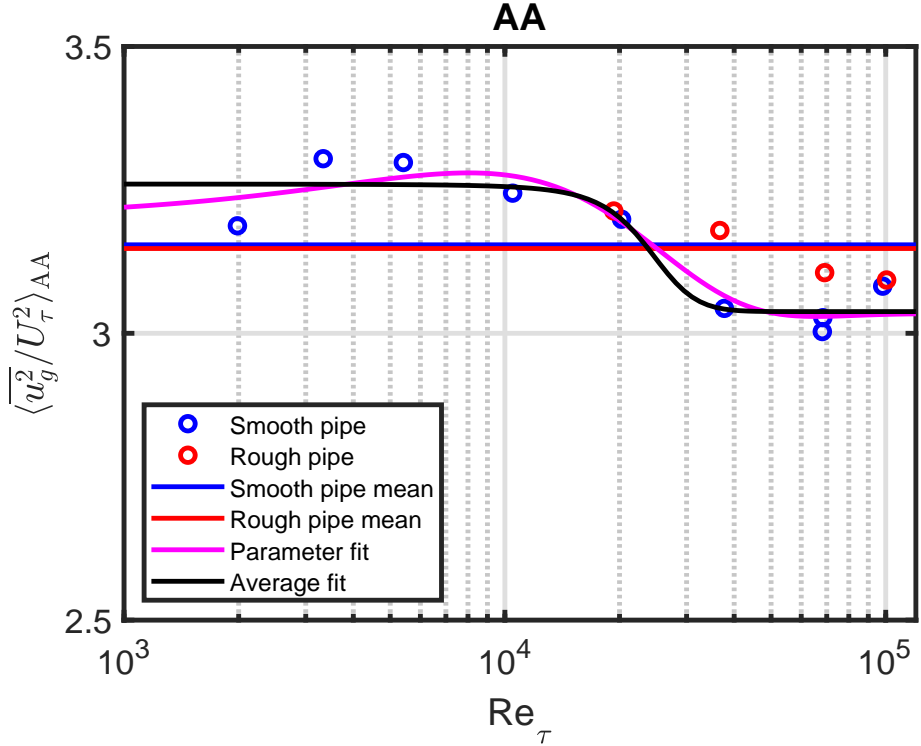}
\caption{AA average as a function of $Re_{\tau}$. Fit to the sum using Equation (\ref{eq:Q_fit}), both for the parameters and the average.}
\label{fig:AA_fits_param_avrg}
\end{figure}

Corresponding tanh fits, both using the individual parameter fits and a fit to the average, are included in Figure \ref{fig:AA_fits_param_avrg}. Fits for the averages can be found in Table \ref{tab:tanh_fit_averages}. The transitional Reynolds numbers are in the range 22000-26000 which is roughly a factor of two higher than for the individual parameters. We also note that the individual parameter fits decrease for low Reynolds number which can not be captured by the average fits. Comparing $R^2$ from Tables \ref{tab:tanh_fit_parameters} and \ref{tab:tanh_fit_averages}, we conclude that the parameter-based fits are a better match than the average-based fits.

The fits are made to the smooth pipe measurements, and we observe that this fit is not a perfect match for the rough pipe measurements.

Figures illustrating similar results for the AM and VA cases can be found in \ref{sec:app_loc_glob}. We focus on AA here since it will be used for the turbulence intensity (TI) scaling in Section \ref{subsec:TI_scaling}.

\begin{table}[!ht]
\caption{Fits to averages using Equation (\ref{eq:Q_fit}).} % title of Table
\centering % used for centering table
\begin{tabular}{cccccc} % centered columns (4 columns)
\hline\hline %inserts double horizontal lines
Average & $a$ & $b$ & $c$ & $d$ & $R^2$\\  % inserts table
%heading
\hline % inserts single horizontal line
$\biggl \langle \frac{{\overline{u^2_g}}}{U_{\tau}^2} \biggr \rangle_{\rm AM}$ & 2.39 & -0.078 & 1.60e-4 & 22266 & 0.84\\
$\biggl \langle \frac{{\overline{u^2_g}}}{U_{\tau}^2} \biggr \rangle_{\rm AA}$ & 3.15 & -0.11 & 1.40e-4 & 23677 & 0.89\\
$\biggl \langle \frac{{\overline{u^2_g}}}{U_{\tau}^2} \biggr \rangle_{\rm VA}$ & 3.62 & -0.12 & 1.18e-4 & 26398 & 0.87\\
\hline %inserts single line
\end{tabular}
\label{tab:tanh_fit_averages} % is used to refer this table in the text
\end{table}

\section{Discussion}
\label{sec:discussion}

\subsection{Peak scaling}
\label{subsec:peak_scal}

Our introduction of the global peak as a mathematical abstraction is meant to capture turbulence production, both clearly identified as an inner peak and also - more controversially - as an outer peak which might emerge for high Reynolds numbers. It has e.g. been proposed that the outer peak is consequence of an invalid use of Taylor's "frozen turbulence" hypothesis \cite{taylor_a}. Here, a single convection velocity is assumed for all eddy scales at a given point. However, "It is suspected that the larger-scale coherent attached eddies are convected downstream at a faster rate than the smaller-scale coherent eddies" \cite{perry_b}.

We propose that the global peak captures both behaviour of the established inner peak and a corresponding outer peak if it exists: The inner (outer) peak dominates for lower (higher) Reynolds numbers, respectively. One issue is that for high Reynolds numbers, the inner peak is not captured which leads to a risk that the measurements become biased towards the pipe axis.

We will show results in Figures \ref{fig:peak_scaling_lo_Re} and \ref{fig:peak_scaling_hi_Re}. The only difference between the figures is the maximum Reynolds number, which is $1.2 \times 10^5$ and $10^8$, respectively.

\subsubsection{Peak position}

The normalized distance from the wall $z^+$ for the various peak definitions is shown in the left-hand plots of Figures \ref{fig:peak_scaling_lo_Re} and \ref{fig:peak_scaling_hi_Re}.

Two values which are independent of $Re_{\tau}$, as defined in Equations (\ref{eq:eleven}) and (\ref{eq:fifteen}), are included.

Two additional peak positions which scale with $Re_{\tau}$ are shown, one is the outer peak (or intersection) scaling from \cite{samie_a}:

\begin{equation}
%\label{}
z^+ \rvert_{\rm outer~peak} = 32.66 \times Re_{\tau}^{0.27},
\end{equation}

\noindent and the other is the global peak scaling, which has the asymptotic behaviour:

\begin{eqnarray}
  \lim_{Re_{\tau}\to\infty} z^+ \rvert_{\rm global~peak} &=& \left( \frac{\lim_{Re_{\tau}\to\infty} C_g}{2 \times \lim_{Re_{\tau}\to\infty} A_g} \right)^2 \\
   &=& 1.41 \times 10^{-3} \times Re_{\tau}
\end{eqnarray}

\subsubsection{Peak amplitude}

If we first focus on the inner peak amplitude, earlier results propose either a log-law \cite{samie_a} scaling:.

\begin{equation}
%\label{}
\frac{\overline{u^2}}{U_{\tau}^2} \bigg \rvert_{\rm inner~peak, log-law} = 0.646 \times \log(Re_{\tau}) + 3.54
\end{equation}

\noindent or a power-law scaling \cite{chen_a}:

\begin{equation}
%\label{}
\frac{\overline{u^2}}{U_{\tau}^2} \bigg \rvert_{\rm inner~peak, power-law} = 11.5 - 19.32 \times Re_{\tau}^{-1/4}
\end{equation}

\begin{figure}[!ht]
\centering
\includegraphics[width=6.5cm]{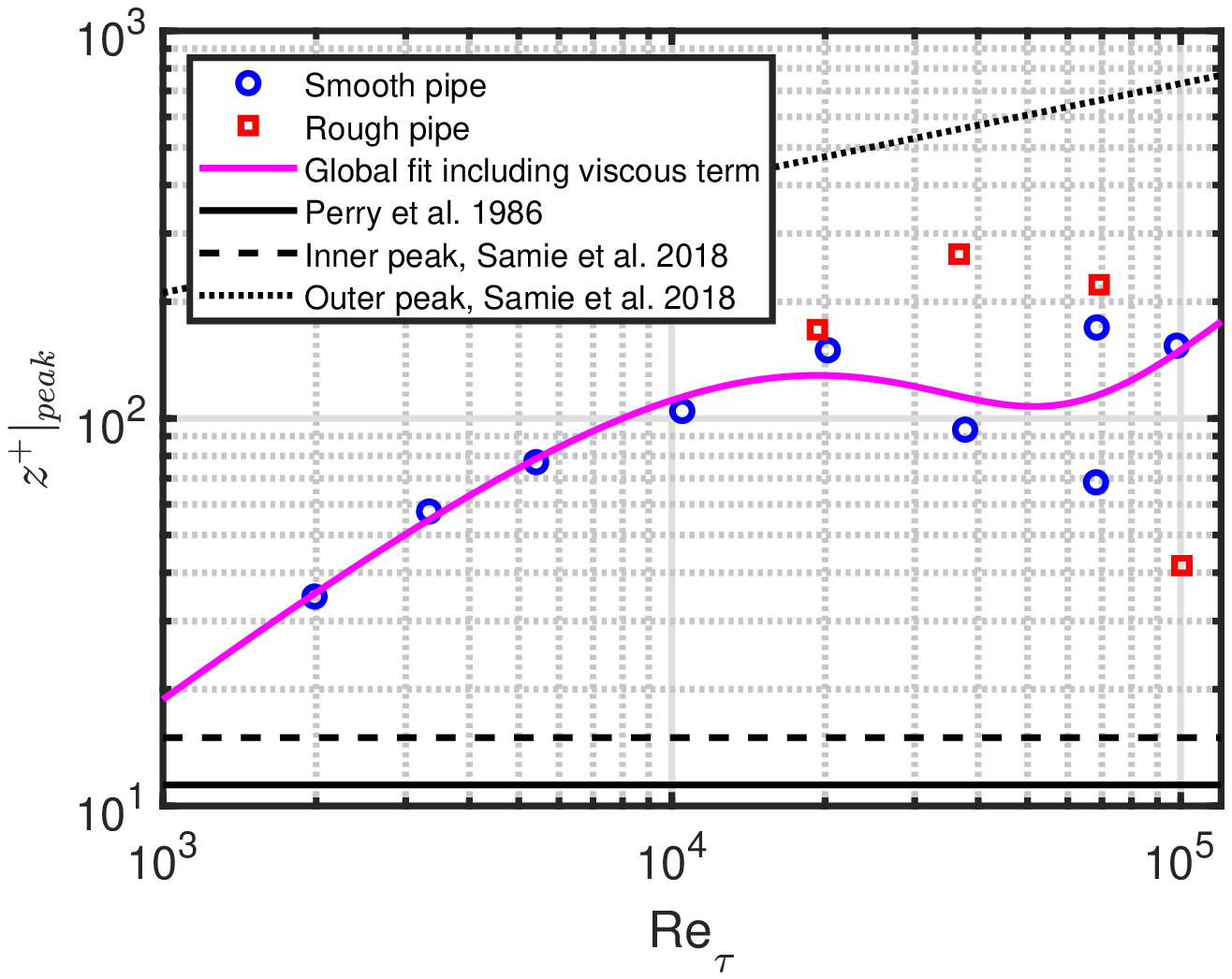}
\hspace{0.3cm}
\includegraphics[width=6.5cm]{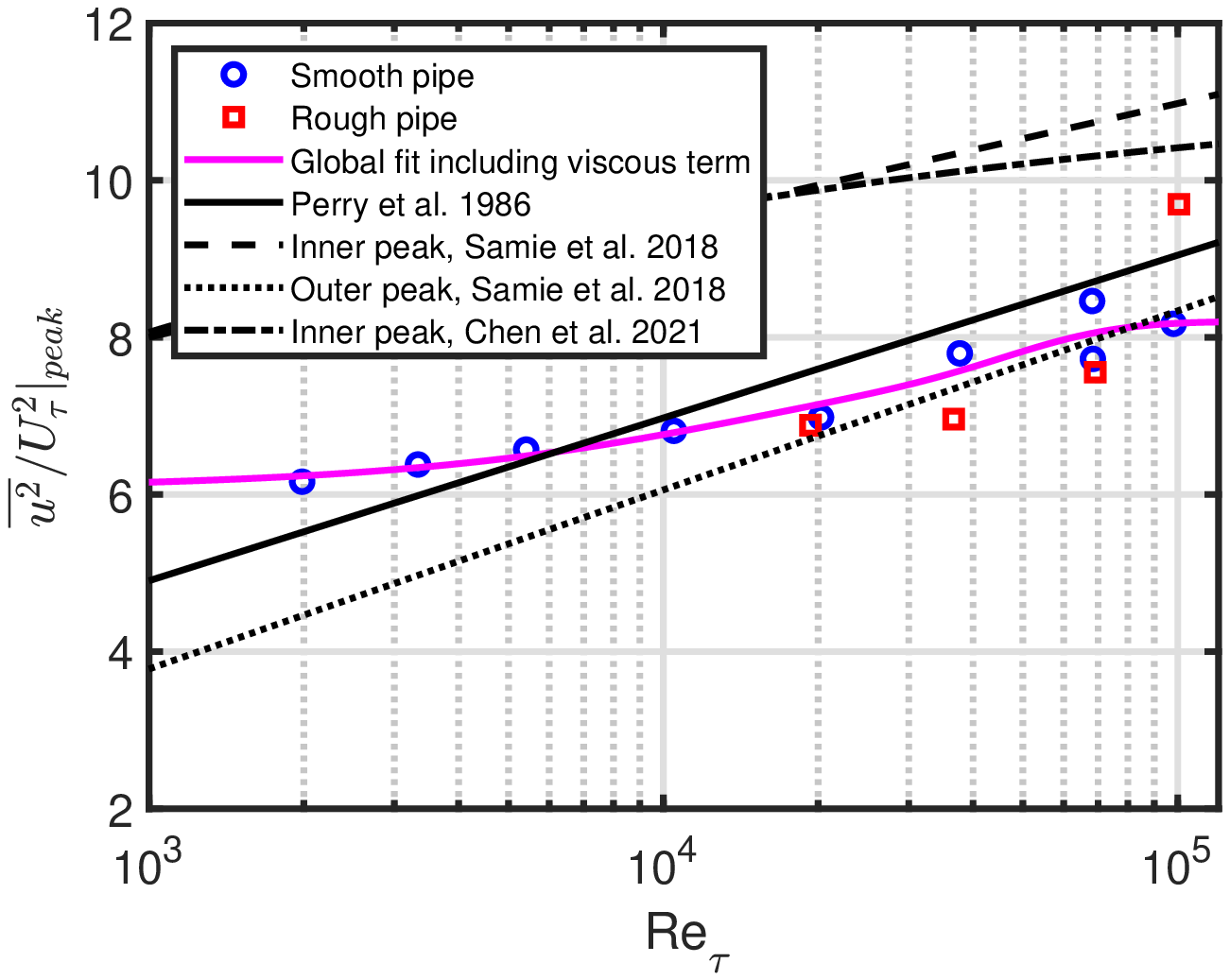}
\caption{Comparison of inner, outer and global peaks. Left-hand plot: $z^+ \rvert_{\rm peak}$ as a function of $Re_{\tau}$, right-hand plot: $\frac{{\overline{u^2_g}}}{U_{\tau}^2} \bigg \rvert_{\rm peak}$ as a function of $Re_{\tau}$.}
\label{fig:peak_scaling_lo_Re}
\end{figure}

\begin{figure}[!ht]
\centering
\includegraphics[width=6.5cm]{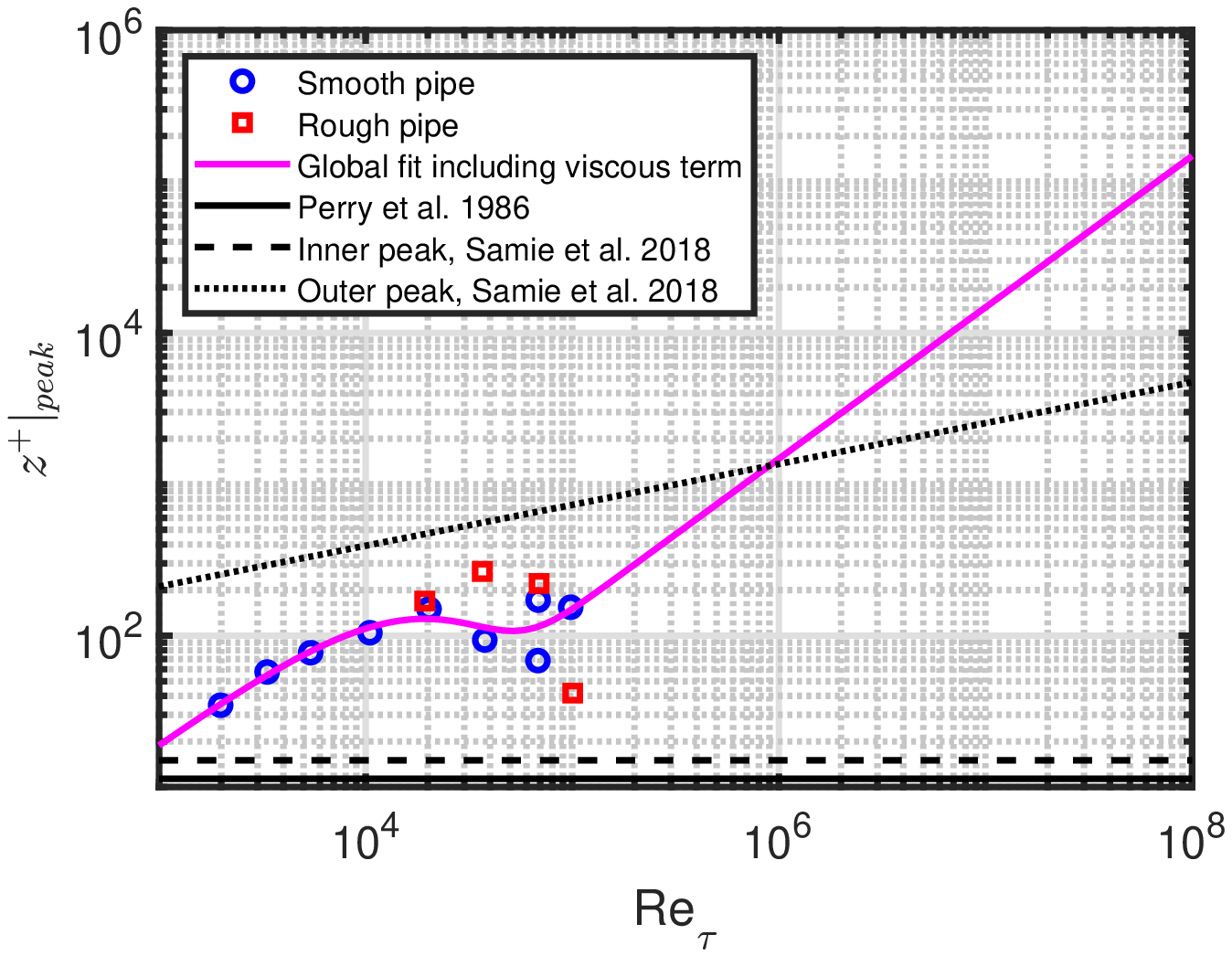}
\hspace{0.3cm}
\includegraphics[width=6.5cm]{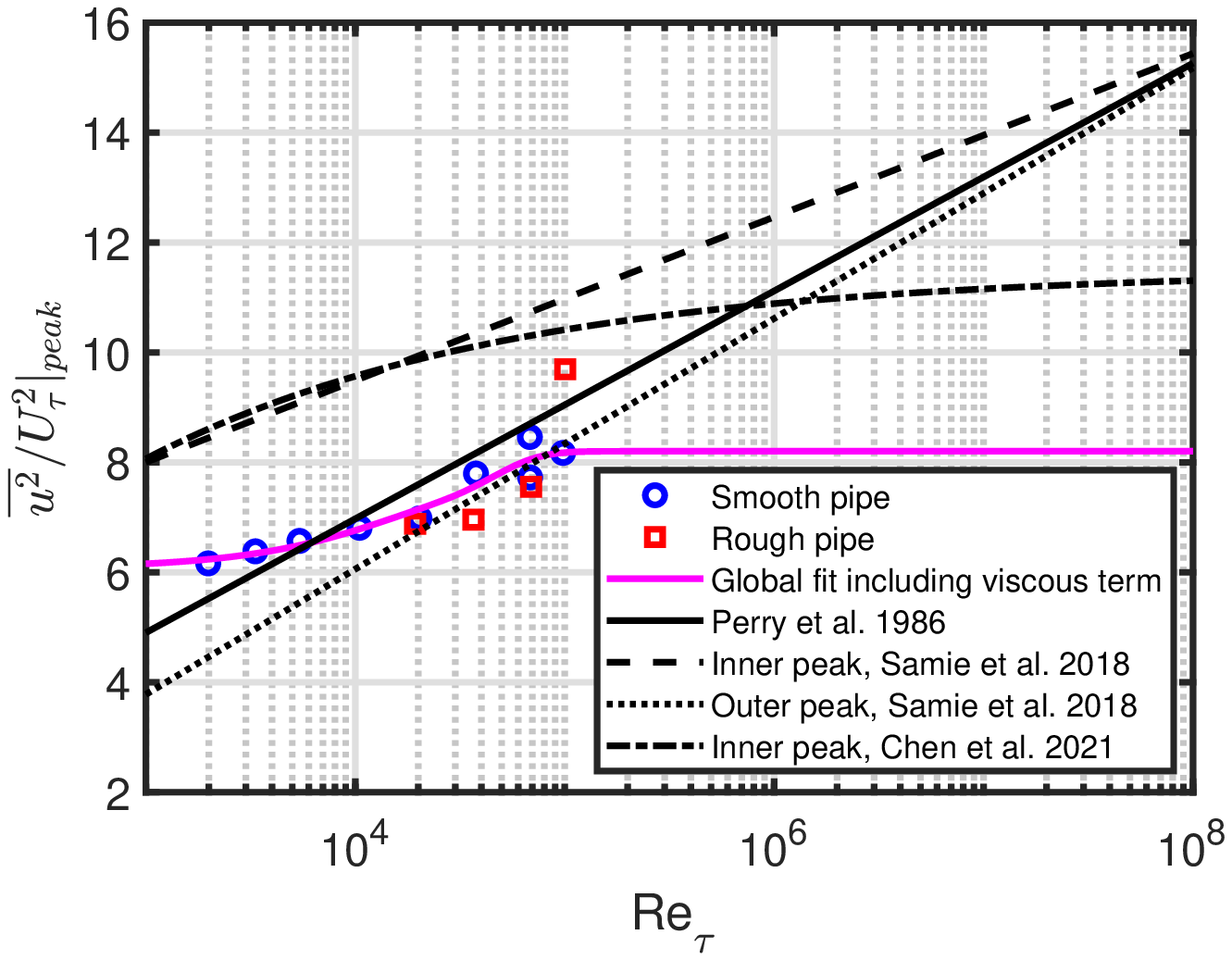}
\caption{Comparison of inner, outer and global peaks, extended $Re_{\tau}$-scale. Left-hand plot: $z^+ \rvert_{\rm peak}$ as a function of $Re_{\tau}$, right-hand plot: $\frac{{\overline{u^2_g}}}{U_{\tau}^2} \bigg \rvert_{\rm peak}$ as a function of $Re_{\tau}$.}
\label{fig:peak_scaling_hi_Re}
\end{figure}

Both scalings are shown in the right-hand plots of Figures \ref{fig:peak_scaling_lo_Re} and \ref{fig:peak_scaling_hi_Re} and deviate visibly above $Re_{\tau} \sim 10000$ \cite{basse_a}. Global peak scalings from Equation (\ref{eq:peak_amp_perry}) and our global fit:

\begin{equation}
%\label{}
\frac{{\overline{u^2_g}}}{U_{\tau}^2} \bigg \rvert_{\rm global~peak} = B_{g}-2A_{g} \times (1+\log (C_g)-\log(2A_{g})) + A_{g} \log (Re_{\tau})
\end{equation}

\noindent are also included along with the outer peak (or intersection) scaling from \cite{samie_a}:

\begin{equation}
\label{eq:samie_outer}
\frac{\overline{u^2}}{U_{\tau}^2} \bigg \rvert_{\rm outer~peak, log-law} = 0.99 \times \log(Re_{\tau}) -3.06,
\end{equation}

\noindent which is quite similar to Equation (\ref{eq:peak_amp_perry}) for high Reynolds number where the constant becomes negligible.

It is interesting to note that the asymptotic value of our global fit is a constant:

\begin{equation}
\label{eq:global_amp_lim}
\lim_{Re_{\tau}\to\infty} \frac{{\overline{u^2_g}}}{U_{\tau}^2} \bigg \rvert_{\rm global~peak} = 8.20
\end{equation}

Thus, the position of the global peak increases without bound, but the corresponding amplitude is bounded. This is in contrast to the log-law behaviour of Equations (\ref{eq:peak_amp_perry}) and (\ref{eq:samie_outer}), which deviate from Equation (\ref{eq:global_amp_lim}) for Reynolds numbers beyond $10^5$. It is not possible to evaluate these differences at present; higher Reynolds number measurements are needed.

\subsection{Length scales}

\begin{figure}[!ht]
\centering
\includegraphics[width=6.5cm]{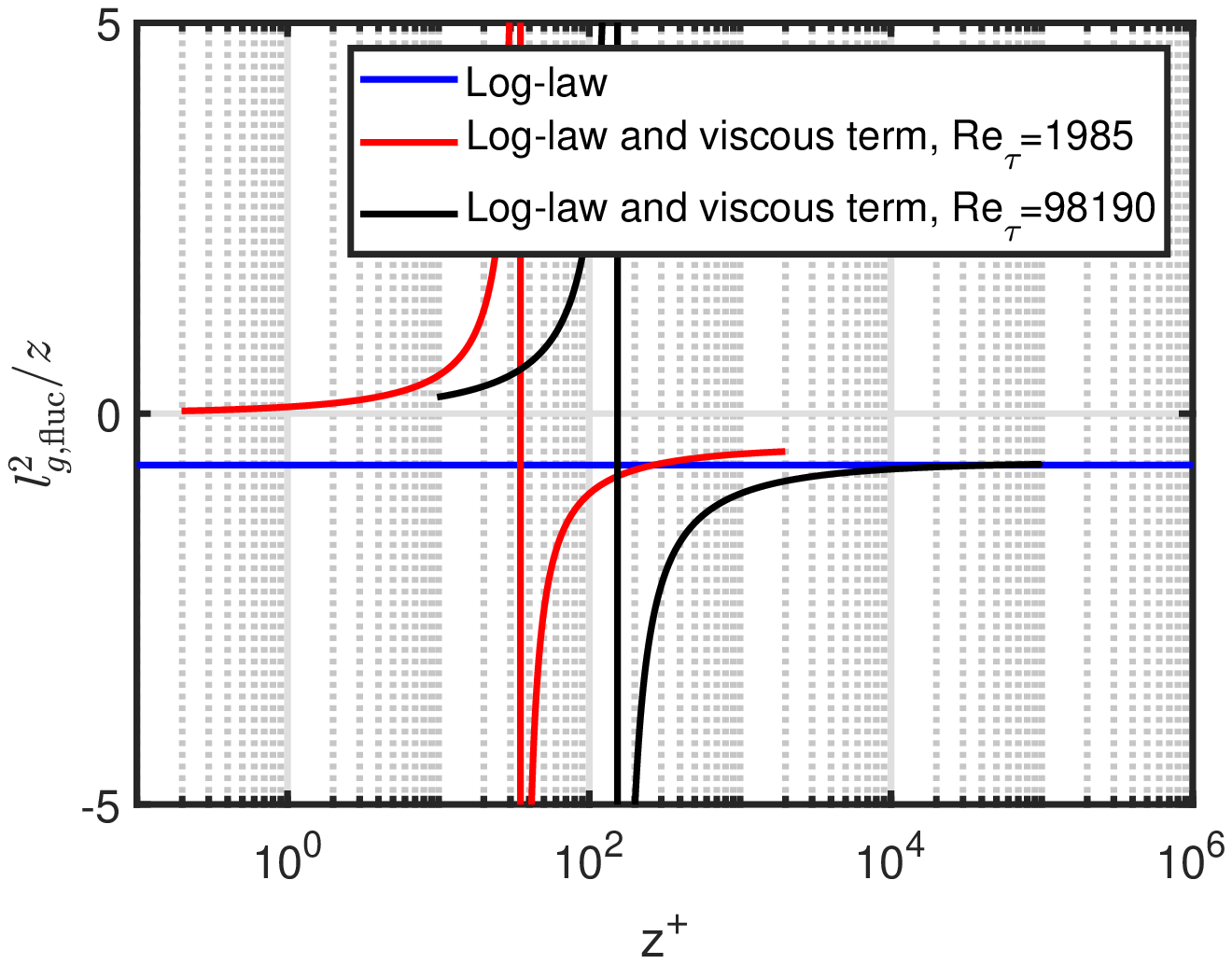}
\hspace{0.3cm}
\includegraphics[width=6.5cm]{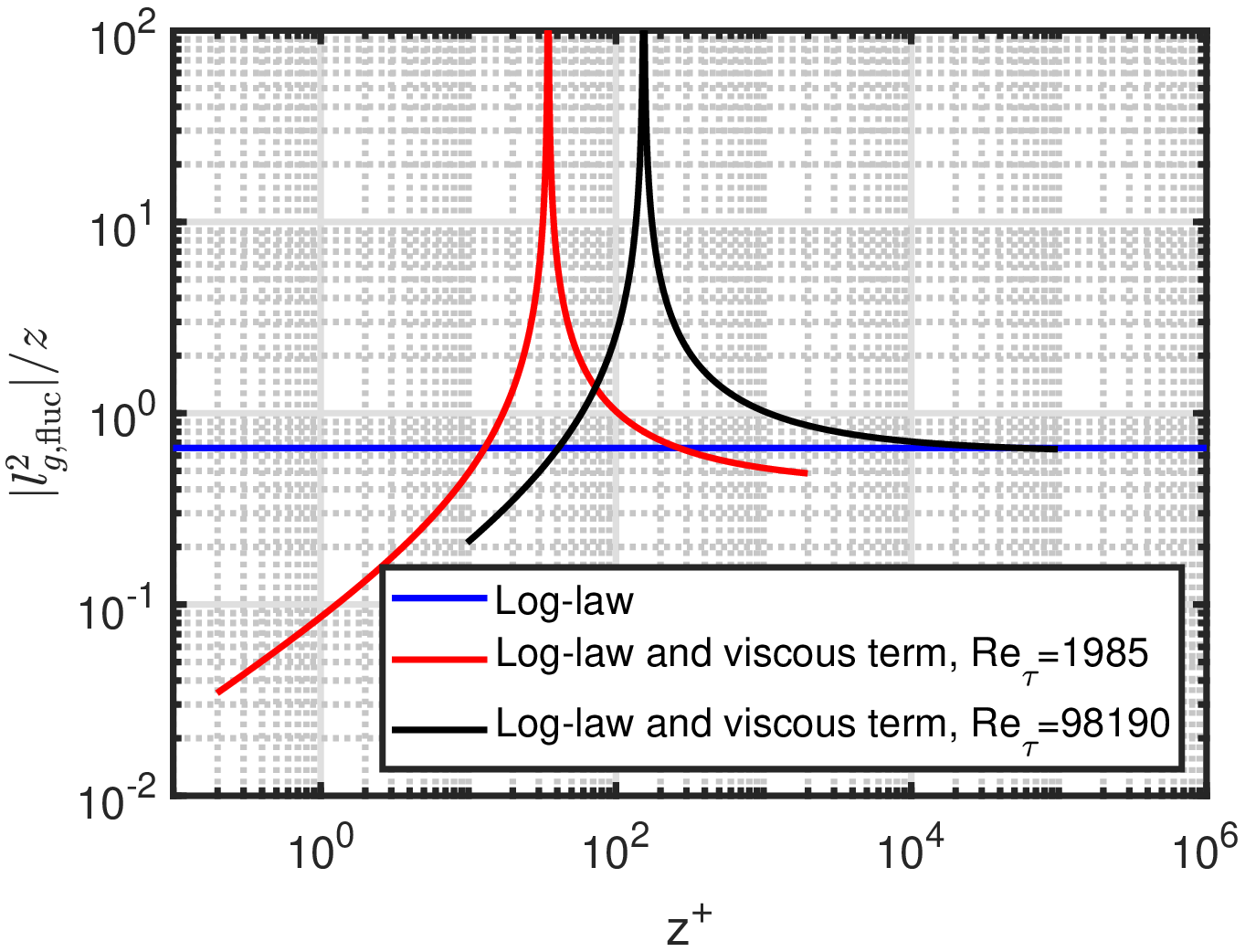}
\caption{The global length scale of the square of the normalised fluctuating velocity divided by $z$ as a function of $z^+$. Left-hand plot: $l_{g{\rm ,fluc}}^2/z$, right-hand plot: $|l_{g{\rm ,fluc}}^2|/z$. The horizontal log-law line is defined using $A_g$ from \cite{basse_a} (without the viscous term): (-)1/1.52.}
\label{fig:length_scales}
\end{figure}

A global length scale for the square of the normalised fluctuating velocity can be constructed by spatially differentiating Equation (\ref{eq:fluc_sq_perry}):

\begin{equation}
%\label{}
l_{g{\rm ,fluc}}^2 = z \times \frac{2 \sqrt{z/\delta}}{C_g/\sqrt{Re_{\tau}}-2A_g\sqrt{z/\delta}} = z\times \frac{2\sqrt{z^+}}{C_g-2A_g\sqrt{z^+}},
\end{equation}

\noindent which can be rewritten as:

\begin{equation}
%\label{}
|l_{g{\rm ,fluc}}^2|/z = \bigg \vert \frac{2 \sqrt{z/\delta}}{C_g/\sqrt{Re_{\tau}}-2A_g\sqrt{z/\delta}} \bigg \vert = \bigg \vert \frac{2\sqrt{z^+}}{C_g-2A_g\sqrt{z^+}} \bigg \vert,
\end{equation}

\noindent where we take the absolute value, since the denominator becomes negative for $z^+$ values larger than the peak position; both $l_{g{\rm ,fluc}}^2/z$ and $|l_{g{\rm ,fluc}}^2|/z$ are shown in Figure \ref{fig:length_scales}. The limits towards the wall and towards the pipe axis are:

\begin{equation}
%\label{}
  \lim_{z^+\to0} l_{g{\rm ,fluc}}^2/z = \frac{2}{C_g} \times \sqrt{z^+}
\end{equation}

\noindent and:

\begin{equation}
%\label{}
  \lim_{z^+\to\infty} l_{g{\rm ,fluc}}^2/z = -\frac{1}{A_g},
\end{equation}

\noindent so the global length scales as $z\sqrt{z^+}$ towards the wall and as $z$ towards the pipe axis. Thus, the scaling close to the wall is mixed scaling, i.e. a combination of inner ($z^+$) and outer ($z/\delta$) coordinates \cite{degraaff_a}. Interpreting this in terms of active and inactive vortex motion \cite{townsend_a}, we would propose that both active and inactive motions are important close to the wall, but that inactive vortex motion dominates towards the pipe axis.

From the log-law for the mean velocity, we have a corresponding length scale:

\begin{equation}
%\label{}
l_{g{\rm ,mean}}/z = \kappa_g,
\end{equation}

\noindent where $\kappa_g$ is the global von K\'arm\'an constant which approaches 0.34 for high Reynolds numbers \cite{basse_a}.

\subsection{Asymptotic behaviour of the viscous term}

Since the asymptotic value of $C_g/\sqrt{Re_{\tau}}$ is finite and non-zero, viscosity - or a similar effect - remains of importance. This finite value also leads to the fact that the global peak amplitude approaches a constant value. It is unclear whether this is a viscosity effect or another physical phenomenon, e.g. a vortex effect \cite{eyink_a}: Here, it is found that "anomalous energy dissipation" is only found for pipe flow if the walls are rough. However, we do not see a clear distinction of the viscous term when we compare the smooth and rough pipe results.

Our research supports the finding that the viscous effect exists for both smooth and rough pipes. This is in line with the fact that we have previously shown that the TI scales with friction factor, which has a finite value for both smooth and rough pipes.

\subsection{Davidson-Krogstad model}

An alternative log-law model for streamwise velocity fluctuations which does not interpret the results as attached eddies has been proposed by Davidson and Krogstad (DK) \cite{davidson_a} . The resulting equation for the streamwise fluctuations in the log-law region has a structure similar to Equation (\ref{eq:fluc_sq_perry}) with:

\begin{eqnarray}
  A_{l,{\rm DK}} &=& 0.91 \\
  B_{l,{\rm DK}} &=& 1.36-0.91 \times \log(P/\varepsilon) \\
  C_{l,{\rm DK}} &=& 3.88 \times (P/\varepsilon)^{-1/2},
\end{eqnarray}

\noindent where $P$ is the rate of turbulent energy production and $\varepsilon$ is the energy dissipation rate. Thus, we can estimate a global average of $\langle P/\varepsilon \rangle$ by comparing our global fit to the local DK model:

\begin{eqnarray}
  B_g &=& 1.36-0.91 \times \log\langle P/\varepsilon \rangle \\
  \langle P/\varepsilon \rangle &=& \exp (1.49-B_g/0.91)
\end{eqnarray}

\begin{figure}[!ht]
\centering
\includegraphics[width=12cm]{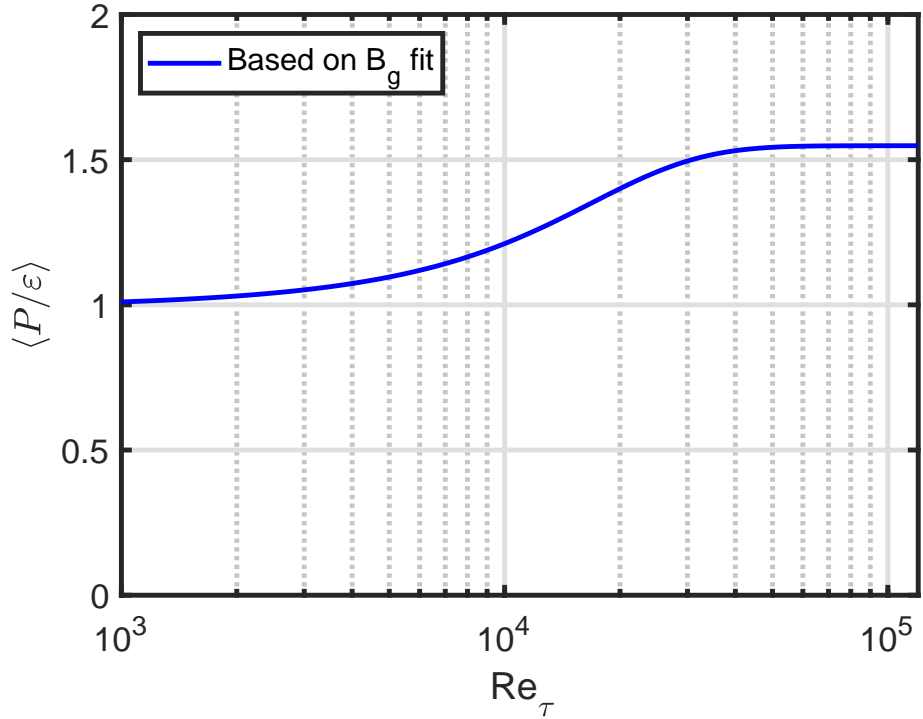}
\caption{Predicted $\langle P/\varepsilon \rangle$ as a function of $Re_{\tau}$.}
\label{fig:P_div_epsilon}
\end{figure}

The result of the comparison can be found in Figure \ref{fig:P_div_epsilon}. The local/global comparison appears to yield reasonable results, where $\langle P/\varepsilon \rangle$ is around one for low Reynolds numbers (production balances dissipation), but rises to a higher plateau for high Reynolds numbers (production is around 1.5 times dissipation).

Using $\langle P/\varepsilon \rangle$ based on $B_g$ for the DK model implies that both $B_{l,{\rm DK}}$ and $C_{l,{\rm DK}}$ decrease with increasing Reynolds number; for the global parameters, we also see a decrease of $B_g$, but an increase for $C_g$. This points to the main differences between local and global behaviour being due to the viscous term.

\vspace{0.5cm}

\subsection{Turbulence intensity scaling with friction factor}
\label{subsec:TI_scaling}

\begin{figure}[!ht]
\centering
\includegraphics[width=6.5cm]{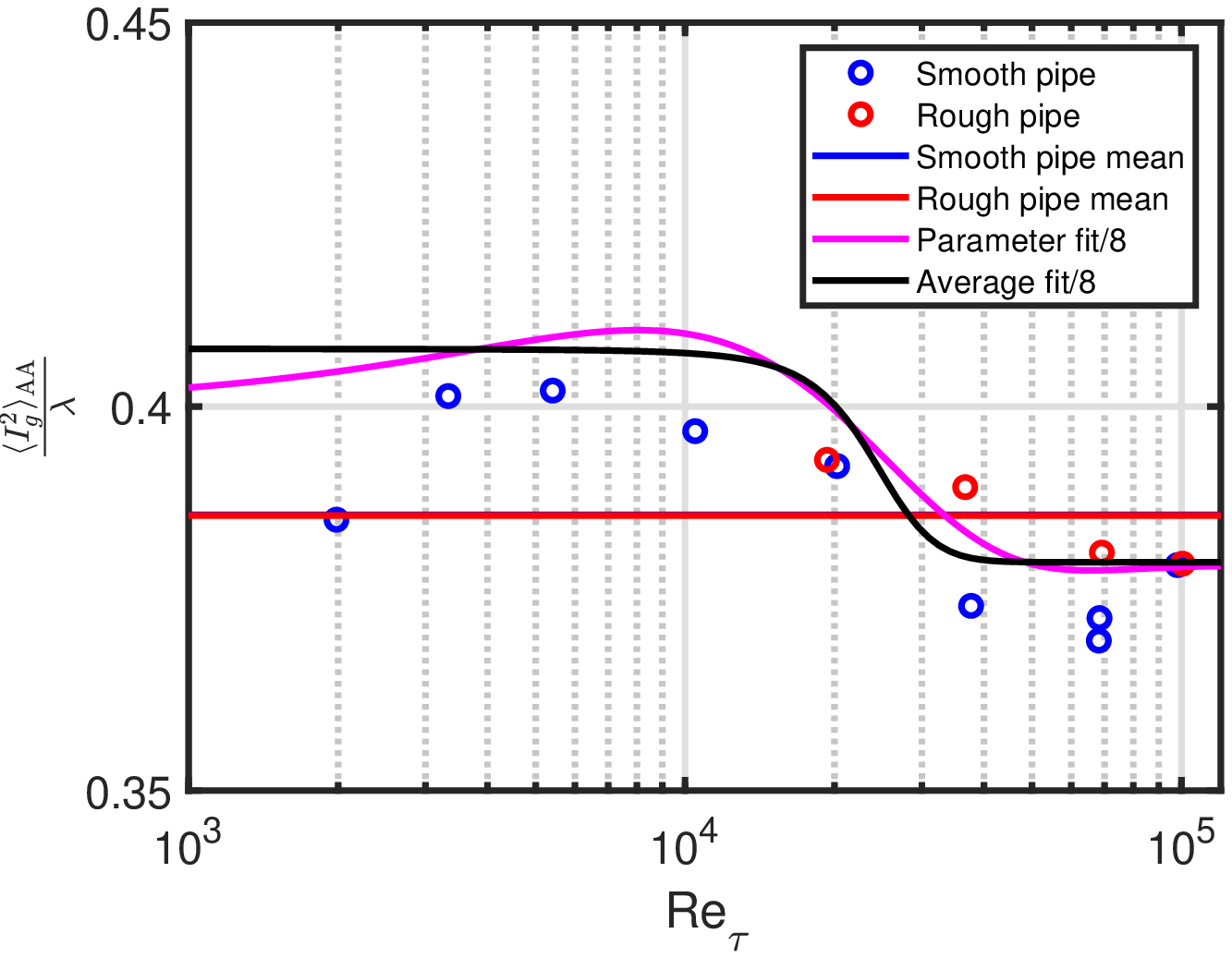}
\hspace{0.3cm}
\includegraphics[width=6.5cm]{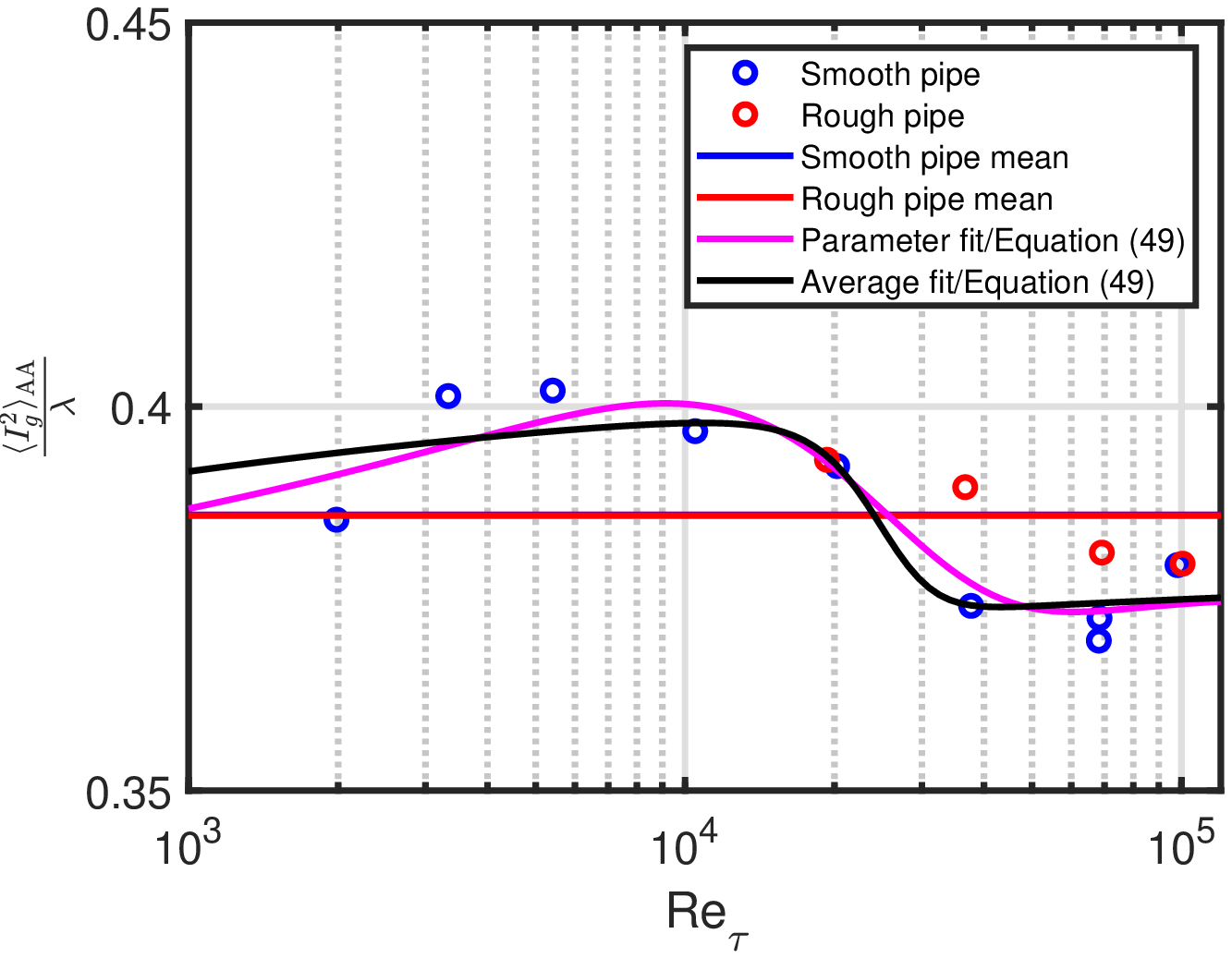}
\caption{$\frac{\langle I_g^2 \rangle_{\rm AA}}{\lambda}$ as a function of $Re_{\tau}$. Mean values of the measurements are shown along with parameter and average fits from the right-hand plot of Figure \ref{fig:fluc_AA_avrg}. Left-hand plot: Fits divided by 8, right-hand plot: Fits divided by Equation (\ref{eq:eight_1par}).}
\label{fig:I_AA_fric_scal}
\end{figure}

We define the TI as $I^2=\overline{u^2}/U^2$, where $U$ is the mean velocity. Our results have an impact on the friction factor scaling of the TI, see the left-hand plot in Figure \ref{fig:I_AA_fric_scal}. Here, we have assumed that:

\begin{equation}
%\label{}
\frac{\langle I_g^2 \rangle_{\rm AA}}{\lambda}   = \frac{1}{\lambda} \times \frac{\biggl \langle \frac{{\overline{u^2_g}}}{U_{\tau}^2} \biggr \rangle_{\rm AA}}{\biggl \langle \frac{U^2_g}{U_{\tau}^2} \biggr \rangle_{\rm AA}}   = \frac{1}{8} \times \biggl \langle \frac{{\overline{u^2_g}}}{U_{\tau}^2} \biggr \rangle_{\rm AA}
\end{equation}

\begin{figure}[!ht]
\centering
\includegraphics[width=12cm]{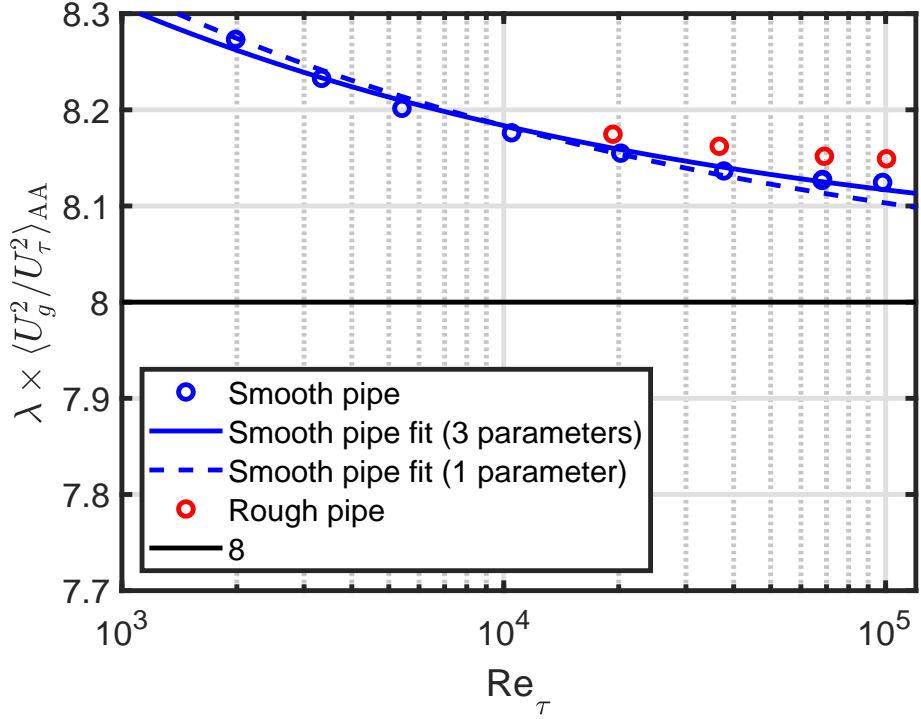}
\caption{Variation of $\lambda \times \biggl \langle \frac{U^2_g}{U_{\tau}^2} \biggr \rangle_{\rm AA}$ as a function of $Re_{\tau}$. The two fits are provided in Equations (\ref{eq:eight_3par}) and (\ref{eq:eight_1par}).}
\label{fig:FF_8}
\end{figure}

However, our result is above the measurements, since the assumption that:

\begin{equation}
%\label{}
\lambda \times \biggl \langle \frac{U^2_g}{U_{\tau}^2} \biggr \rangle_{\rm AA} = 8
\end{equation}

\noindent is not accurate, see Figure \ref{fig:FF_8}. Here, fits are shown using either three parameters:

\begin{equation}
\label{eq:eight_3par}
\lambda \times \biggl \langle \frac{U^2_g}{U_{\tau}^2} \biggr \rangle_{\rm AA} = 8.04 + 1.73 \times Re_{\tau}^{-0.27} \qquad R^2=0.99,
\end{equation}

\noindent or one parameter (assuming a constant term 8 and a fixed exponent of -1/4):

\begin{equation}
\label{eq:eight_1par}
\lambda \times \biggl \langle \frac{U^2_g}{U_{\tau}^2} \biggr \rangle_{\rm AA} = 8 + 1.83 \times Re_{\tau}^{-1/4} \qquad R^2=0.95
\end{equation}

Finally, we can write an expression for the area-averaged TI using Equation (\ref{eq:eight_1par}):

\begin{equation}
  \langle I_g^2 \rangle_{\rm AA} = \frac{\lambda}{8 + 1.83 \times Re_{\tau}^{-1/4}} \times \biggl \langle \frac{{\overline{u^2_g}}}{U_{\tau}^2} \biggr \rangle_{\rm AA},
\end{equation}

\noindent where $\biggl \langle \frac{{\overline{u^2_g}}}{U_{\tau}^2} \biggr \rangle_{\rm AA}$ can be defined using Equation (\ref{eq:Q_fit}) with either parameter fits (Table \ref{tab:tanh_fit_parameters}) or the average fit (Table \ref{tab:tanh_fit_averages}), see the right-hand plot in Figure \ref{fig:I_AA_fric_scal}. Asymptotically, these two alternatives yield:

\begin{equation}
  \lim_{Re_{\tau}\to\infty} \langle I_g^2 \rangle_{\rm AA,parameter~fits} = \frac{3.03}{8} \times \lambda = 0.38 \times \lambda
\end{equation}

\begin{equation}
  \lim_{Re_{\tau}\to\infty} \langle I_g^2 \rangle_{\rm AA,average~fit} = \frac{3.04}{8} \times \lambda = 0.38 \times \lambda,
\end{equation}

\noindent which is very close to what we proposed in \cite{basse_a} for the entire Reynolds number range, i.e. $\langle I_g^2 \rangle_{\rm AA} = 0.39 \times \lambda$.

\section{Conclusions}
\label{sec:conclusions}

We have introduced a global model of the streamwise velocity fluctuations in pipe flow calibrated to Princeton Superpipe measurements. The model includes both a log-law and a viscous term. The global model captures the overall behaviour of the fluctuations but is a physical abstraction with the main purpose of quantifying turbulence intensity scaling with Reynolds number. The model has a single global peak, which is bounded and captures both the inner and outer peaks. In reality, the flow transitions from a lower Reynolds number inner peak dominated flow to a higher Reynolds number outer peak dominated flow.

The parameters can be represented by hyperbolic tangent fits, and exhibit a transition region for $Re_{\tau} \sim 11000-12000$ (parameter fits) or $Re_{\tau} \sim 22000-26000$ (average fits). This is consistent with a simultaneous modification of fluctuation and mean velocities, i.e including a viscous term clarifies that there is indeed a Reynolds number transition for fluctuations as we have previously shown for the mean flow \cite{basse_a}.

The impact of our findings include peak scaling, length scales, the finite non-zero asymptotic value of the viscous term, turbulent energy production/dissipation and turbulence intensity scaling. Finally, we show in \ref{sec:app_wake} that including a wake term does not lead to a clear transition; rather, the model reverts to being similar to a two-parameter power-law.

\paragraph{Acknowledgements}

We thank Professor Alexander J. Smits for making the Princeton Superpipe data publicly available.

\paragraph{Data availability statement}

Data sharing is not applicable to this article as no new data were created or analyzed in this study.

\clearpage

\appendix

\section{Inclusion of the wake term}
\label{sec:app_wake}

As mentioned in Section \ref{sec:local}, we have also carried out our analysis for the case where a wake term is included in addition to the viscous term \cite{marusic_b,marusic_c,kunkel_a}.

\subsection{Model}

Equation (\ref{eq:fluc_sq_perry}) is modified to:

\begin{equation}
%\label{}
\frac{{\overline{u^2_l}}(z)}{U_{\tau}^2} = B_{l} - A_{l} \log (z/\delta) + V(z^+) - W(z/\delta),
\end{equation}

\noindent where

\begin{equation}
\label{eq:fluc_wake}
W(z/\delta)= B_{l} (z/\delta)^2(3-2z/\delta)-A_{l}(z/\delta)^2(1-z/\delta)(1-2z/\delta)
\end{equation}

\noindent is the wake term. Analytical integration of Equation (\ref{eq:fluc_wake}) using Equations (\ref{eq:AM_def})-(\ref{eq:VA_end}) yields these three equations:

\begin{eqnarray}
  \langle W_g \rangle_{\rm AM} &=& \frac{1}{2} \times B_{g{\rm ,wake}} + \frac{1}{60} \times A_{g{\rm ,wake}} \\
  \langle W_g \rangle_{\rm AA} &=& \frac{3}{10} \times B_{g{\rm ,wake}} \\
  \langle W_g \rangle_{\rm VA} &=& \frac{1}{5} \times B_{g{\rm ,wake}} - \frac{1}{140} A_{g{\rm ,wake}}
\end{eqnarray}

Combining these results with the analytical integration only including the viscous term leads to:

\begin{equation}
%\label{}
\biggl \langle \frac{{\overline{u^2_g}}}{U_{\tau}^2} \biggr \rangle_{\rm AM,wake} = \frac{1}{2} \times B_{g{\rm ,wake}} + \frac{59}{60} \times A_{g{\rm ,wake}} - \frac{2 C_{g{\rm ,wake}}}{\sqrt{Re_{\tau}}}
\end{equation}

\begin{equation}
%\label{}
\biggl \langle \frac{{\overline{u^2_g}}}{U_{\tau}^2} \biggr \rangle_{\rm AA,wake} = \frac{7}{10} \times B_{g{\rm ,wake}} + \frac{3}{2} \times A_{g{\rm ,wake}} - \frac{8 C_{g{\rm ,wake}}}{3\sqrt{Re_{\tau}}}
\end{equation}

\begin{equation}
%\label{}
\biggl \langle \frac{{\overline{u^2_g}}}{U_{\tau}^2} \biggr \rangle_{\rm VA,wake} = \frac{4}{5} \times B_{g{\rm ,wake}} + \frac{767}{420} \times A_{g{\rm ,wake}} - \frac{16 C_{g{\rm ,wake}}}{5\sqrt{Re_{\tau}}}
\end{equation}

The solutions to these three equations are:

\begin{equation}
%\label{}
A_{g{\rm ,wake}}= -\frac{1120}{177} \times \biggl \langle \frac{{\overline{u^2_g}}}{U_{\tau}^2} \biggr \rangle_{\rm AM,wake} + \frac{700}{177} \times \biggl \langle \frac{{\overline{u^2_g}}}{U_{\tau}^2} \biggr \rangle_{\rm VA,wake}
\end{equation}

\begin{eqnarray}
  B_{g{\rm ,wake}} &=& -\frac{2200}{531} \times \biggl \langle \frac{{\overline{u^2_g}}}{U_{\tau}^2} \biggr \rangle_{\rm AM,wake} + 30 \times \biggl \langle \frac{{\overline{u^2_g}}}{U_{\tau}^2} \biggr \rangle_{\rm AA,wake} \\
   && - \frac{11900}{531} \times \biggl \langle \frac{{\overline{u^2_g}}}{U_{\tau}^2} \biggr \rangle_{\rm VA,wake}
\end{eqnarray}

\begin{eqnarray}
  C_{g{\rm ,wake}} &=& \sqrt{Re_{\tau}} \\
    && \times \Bigg( \Bigg. -\frac{1645}{354} \times \biggl \langle \frac{{\overline{u^2_g}}}{U_{\tau}^2} \biggr \rangle_{\rm AM,wake} + \frac{15}{2} \times \biggl \langle \frac{{\overline{u^2_g}}}{U_{\tau}^2} \biggr \rangle_{\rm AA,wake} \\
    && - \frac{1295}{354} \times \biggl \langle \frac{{\overline{u^2_g}}}{U_{\tau}^2} \biggr \rangle_{\rm VA,wake} \Bigg. \Bigg)  \nonumber
\end{eqnarray}

\subsection{Parameter fits}

\begin{figure}[!ht]
\centering
\includegraphics[width=6.5cm]{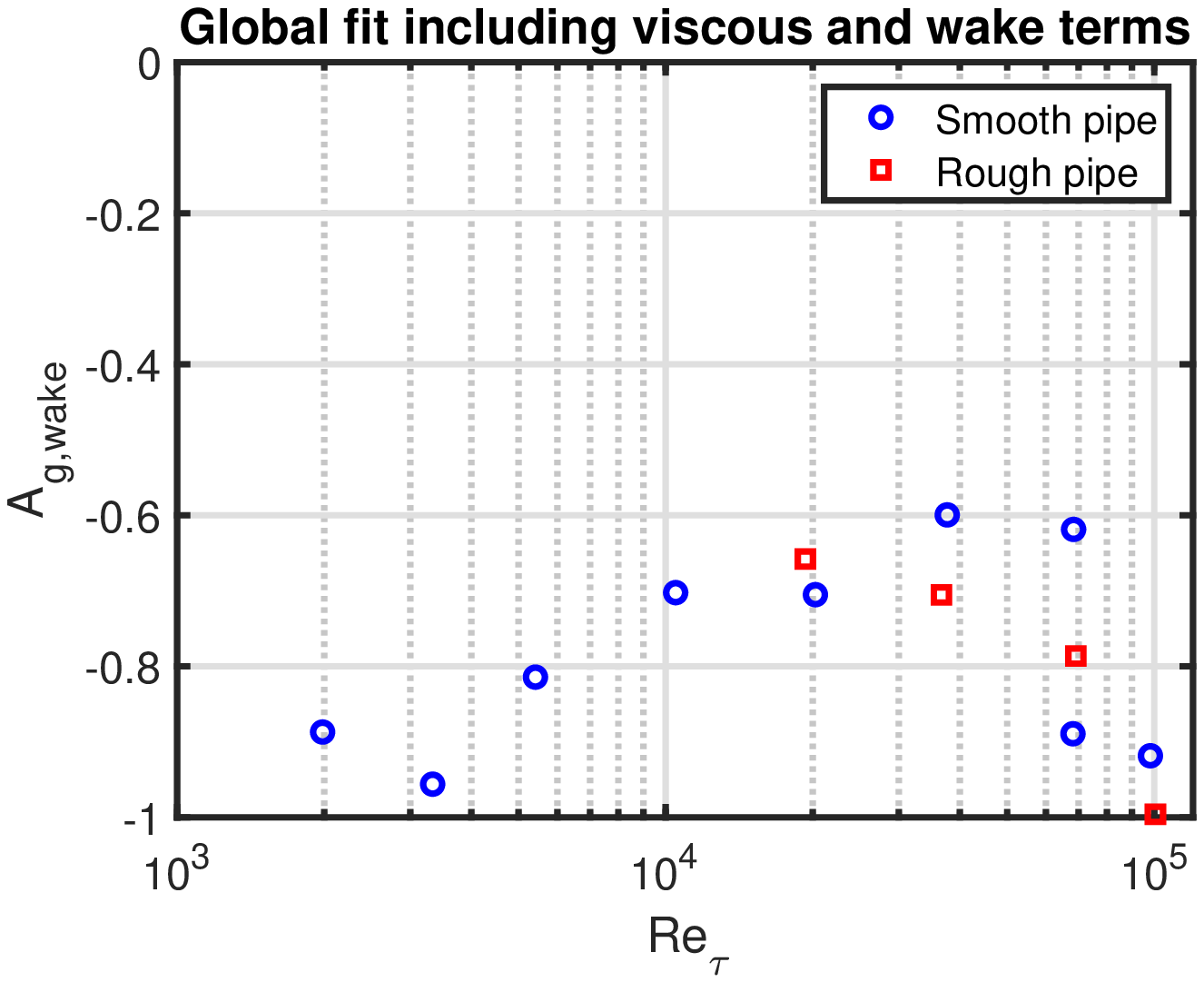}
\hspace{0.3cm}
\includegraphics[width=6.5cm]{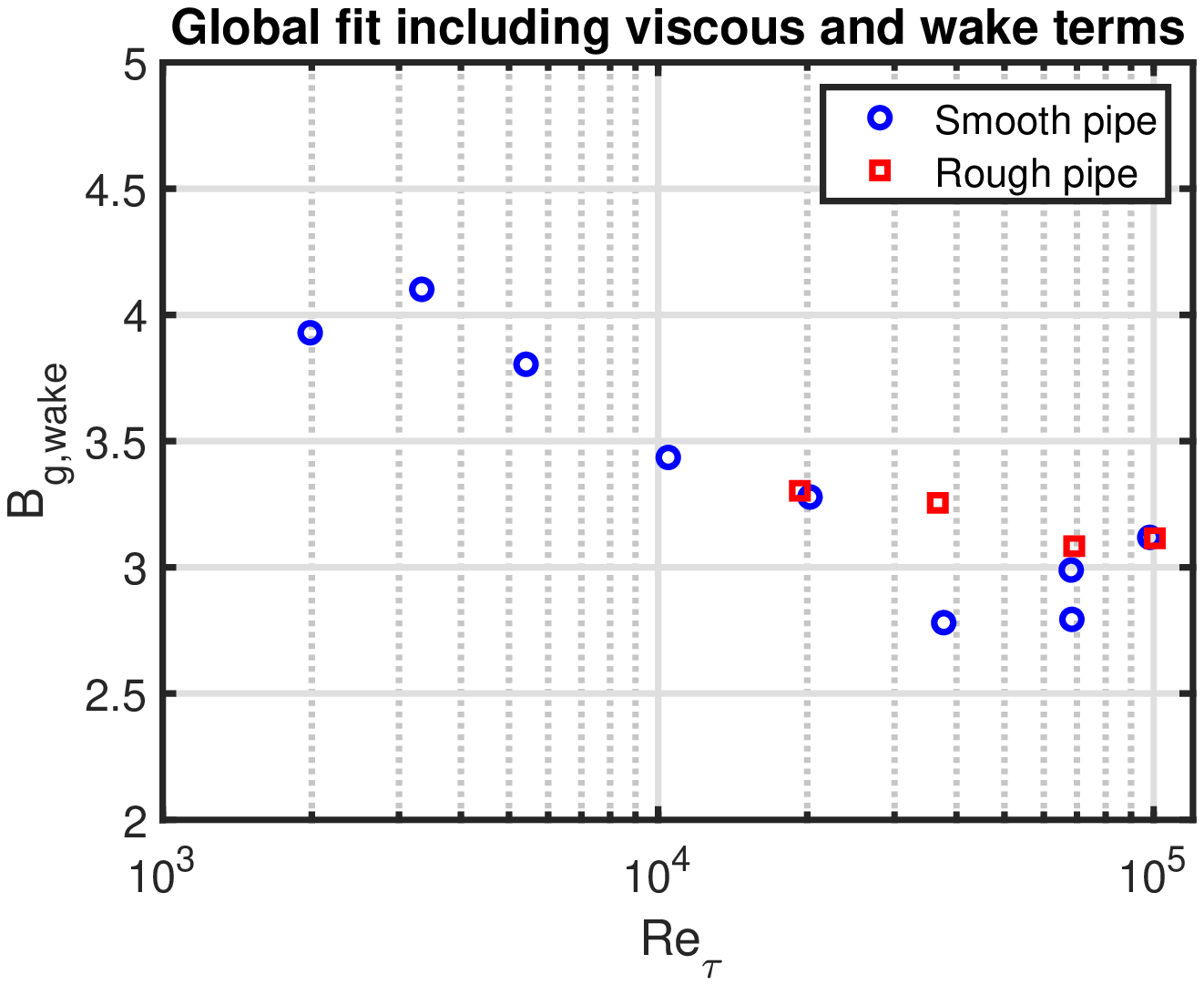}
\caption{Left-hand plot: $A_{g{\rm ,wake}}$ as a function of $Re_{\tau}$, right-hand plot: $B_{g{\rm ,wake}}$ as a function of $Re_{\tau}$.}
\label{fig:glob_wake_fit_A_g}
\end{figure}

\begin{figure}[!ht]
\centering
\includegraphics[width=6.5cm]{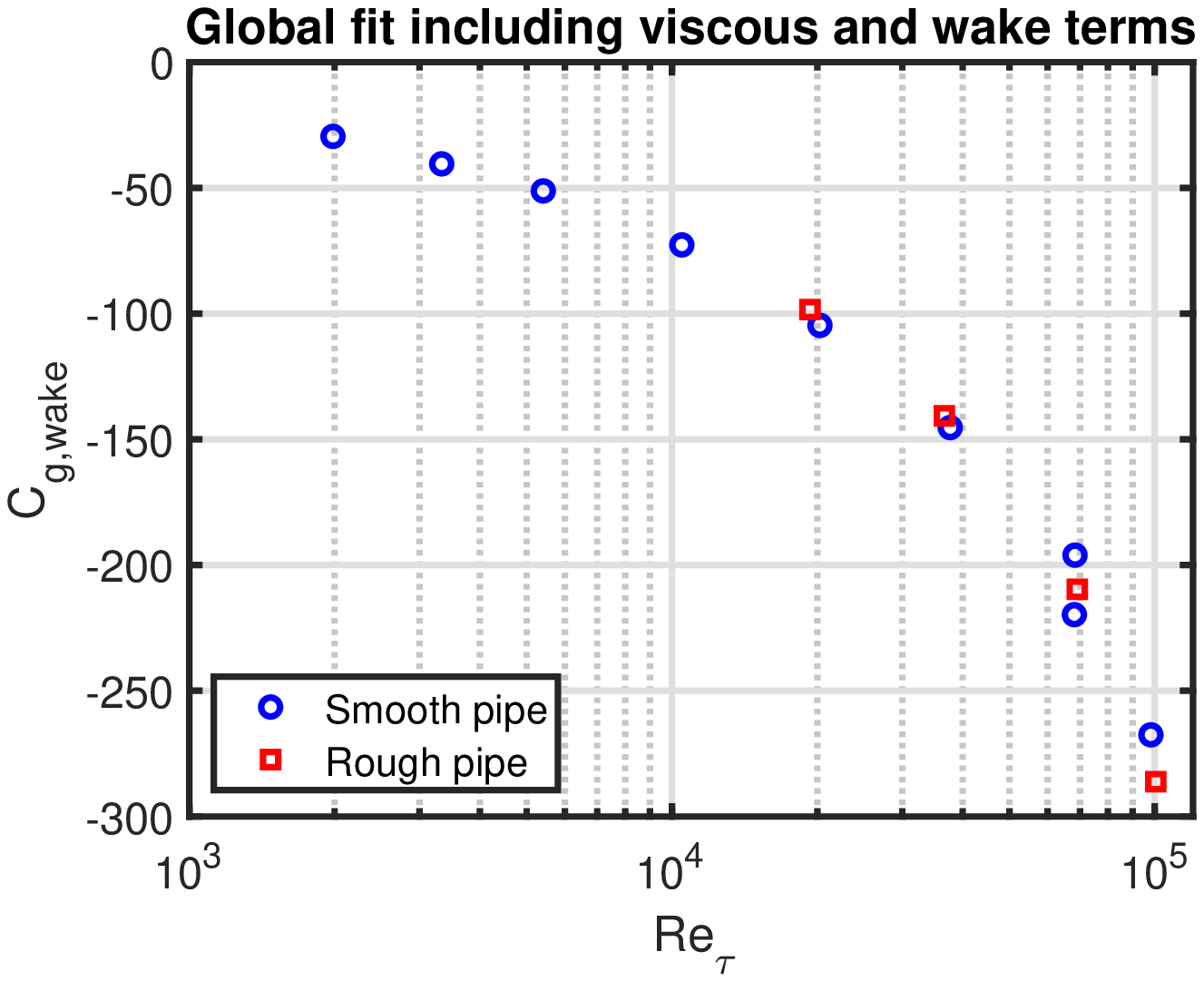}
\hspace{0.3cm}
\includegraphics[width=6.5cm]{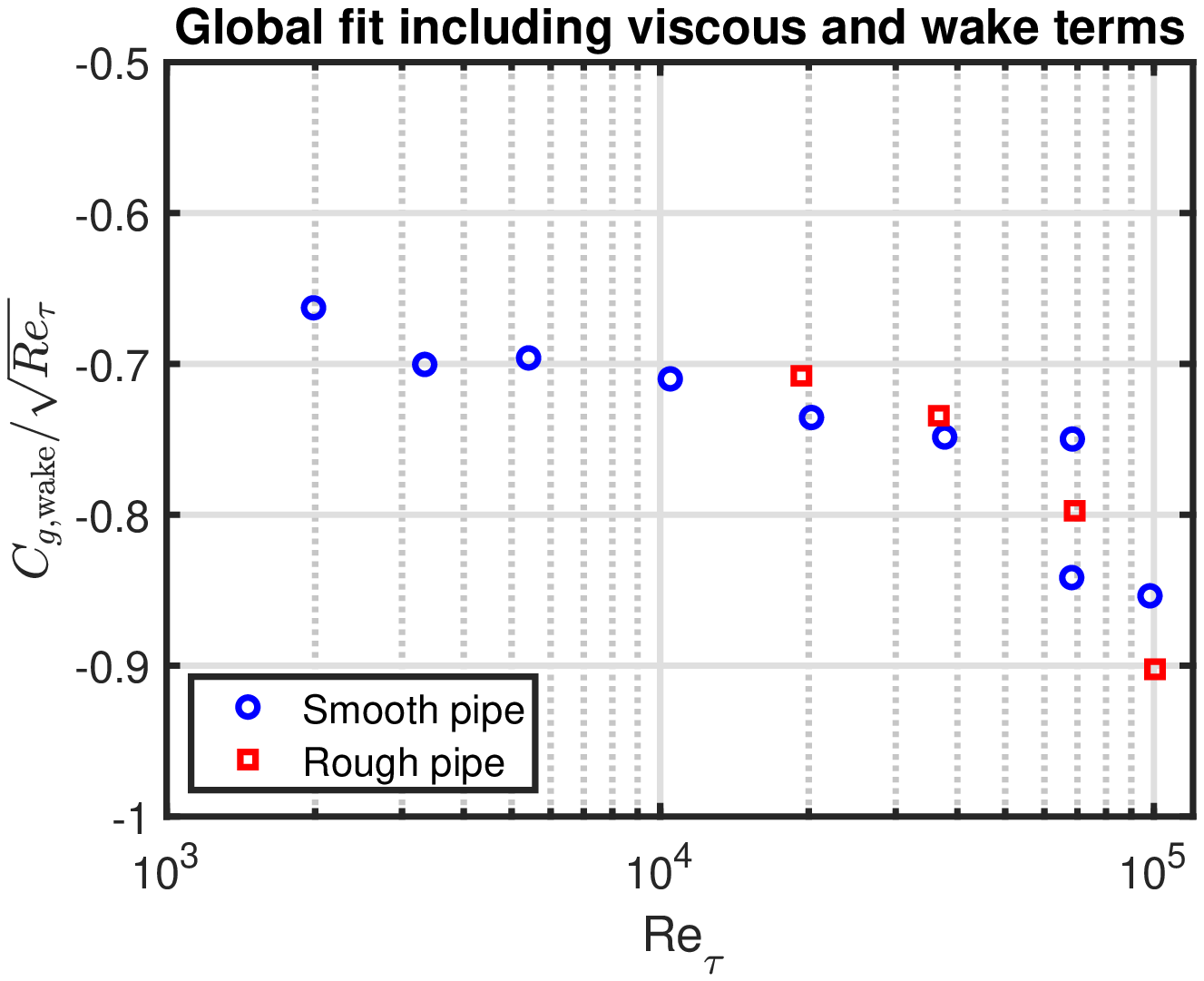}
\caption{Left-hand plot: $C_{g{\rm ,wake}}$ as a function of $Re_{\tau}$, right-hand plot: $C_{g{\rm ,wake}}/\sqrt{Re_{\tau}}$ as a function of $Re_{\tau}$.}
\label{fig:glob_wake_fit_C_g}
\end{figure}

The fits to the wake parameters are shown in Figures \ref{fig:glob_wake_fit_A_g} and \ref{fig:glob_wake_fit_C_g}. $A_{g{\rm ,wake}}$ is negative, which means that the slope of the log-law changes sign. For both $A_{g{\rm ,wake}}$ and $B_{g{\rm ,wake}}$, no strong variation with $Re_{\tau}$ is observed. In contrast, $C_{g{\rm ,wake}}$ has a strong scaling with $Re_{\tau}$. In addition it is negative, so the viscous term becomes positive which is unphysical. The result is an equation with the functional form very similar to Equation (\ref{eq:pow_m0p48}).

\clearpage

\section{Local and global average fits}
\label{sec:app_loc_glob}

For reference, average fits for the AM and VA are collected here.

\begin{figure}[!ht]
\centering
\includegraphics[width=6.5cm]{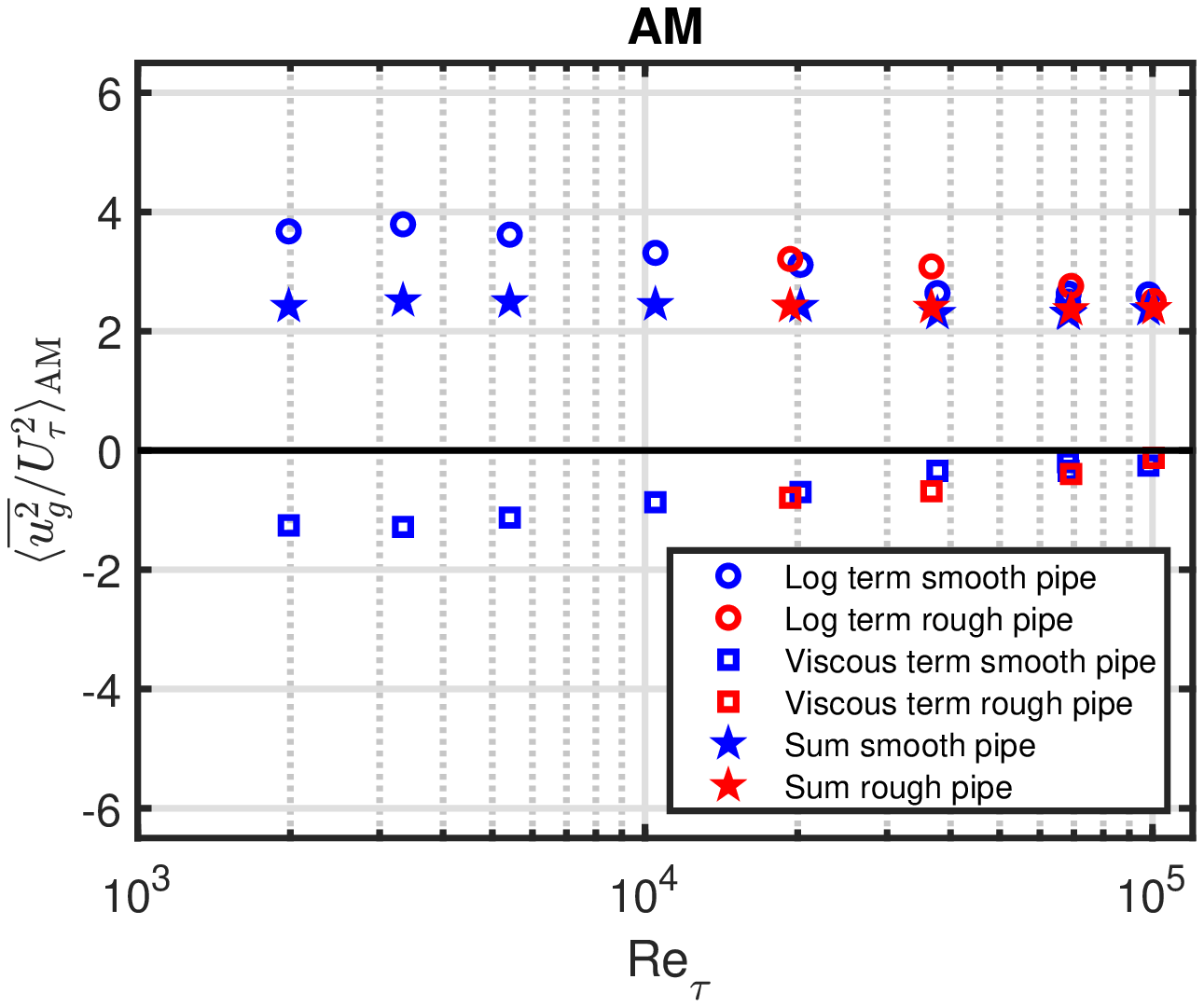}
\hspace{0.3cm}
\includegraphics[width=6.5cm]{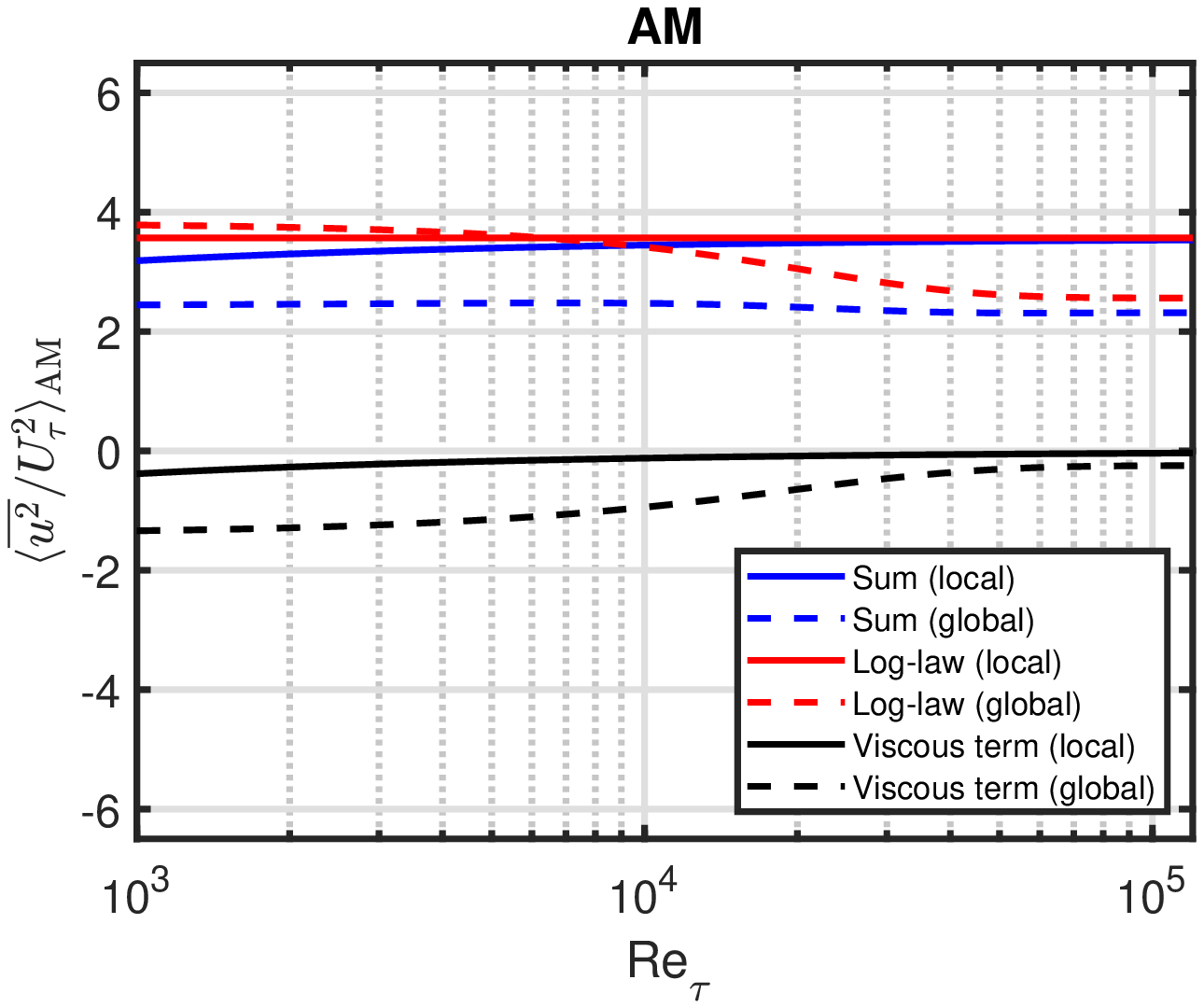}
\caption{Decomposition of the AM average as a function of $Re_{\tau}$. Left-hand plot: Individual points, right-hand plot: Comparison of local and global fits.}
\label{fig:fluc_AM_avrg}
\end{figure}

\begin{figure}[!ht]
\centering
\includegraphics[width=6.5cm]{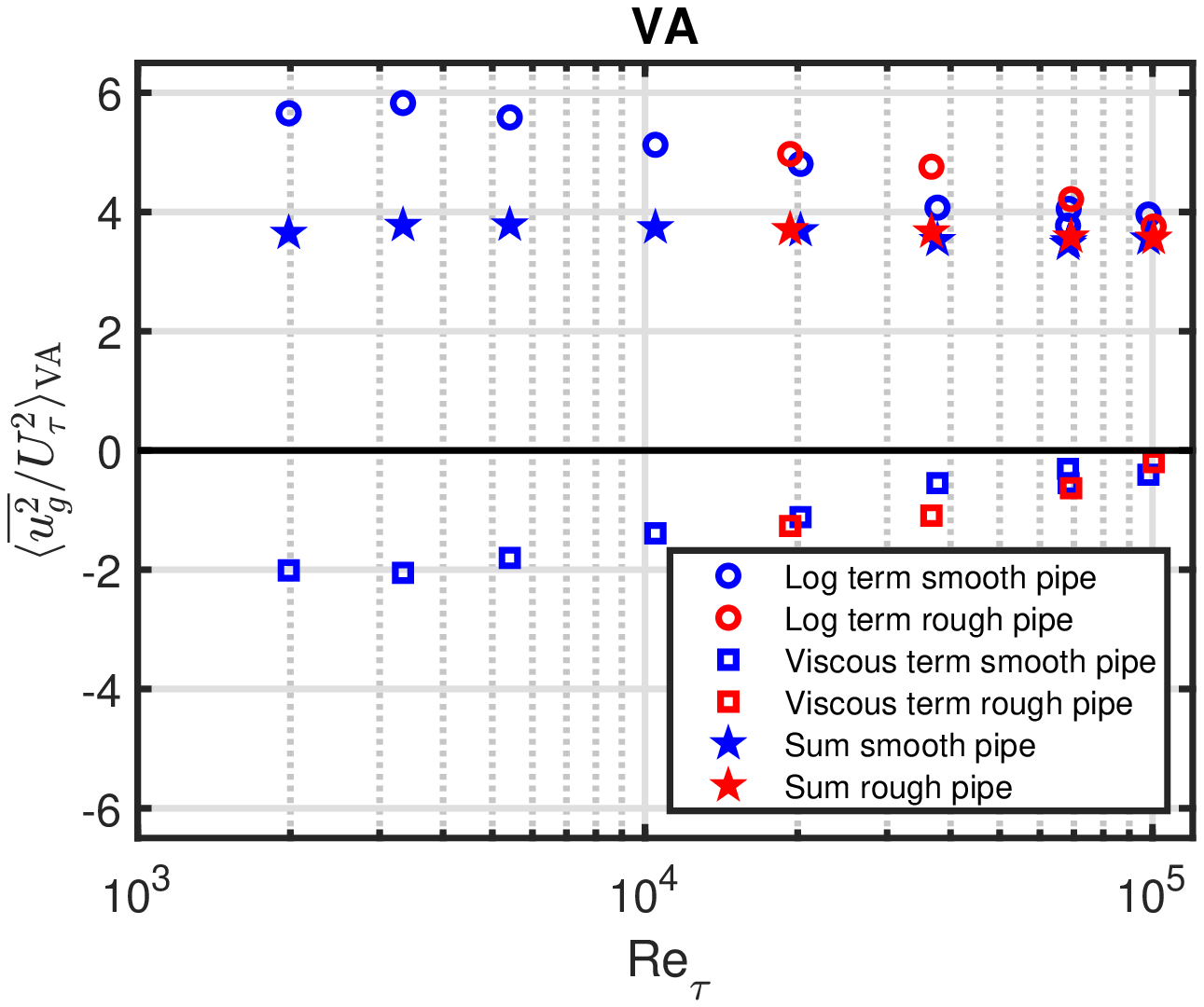}
\hspace{0.3cm}
\includegraphics[width=6.5cm]{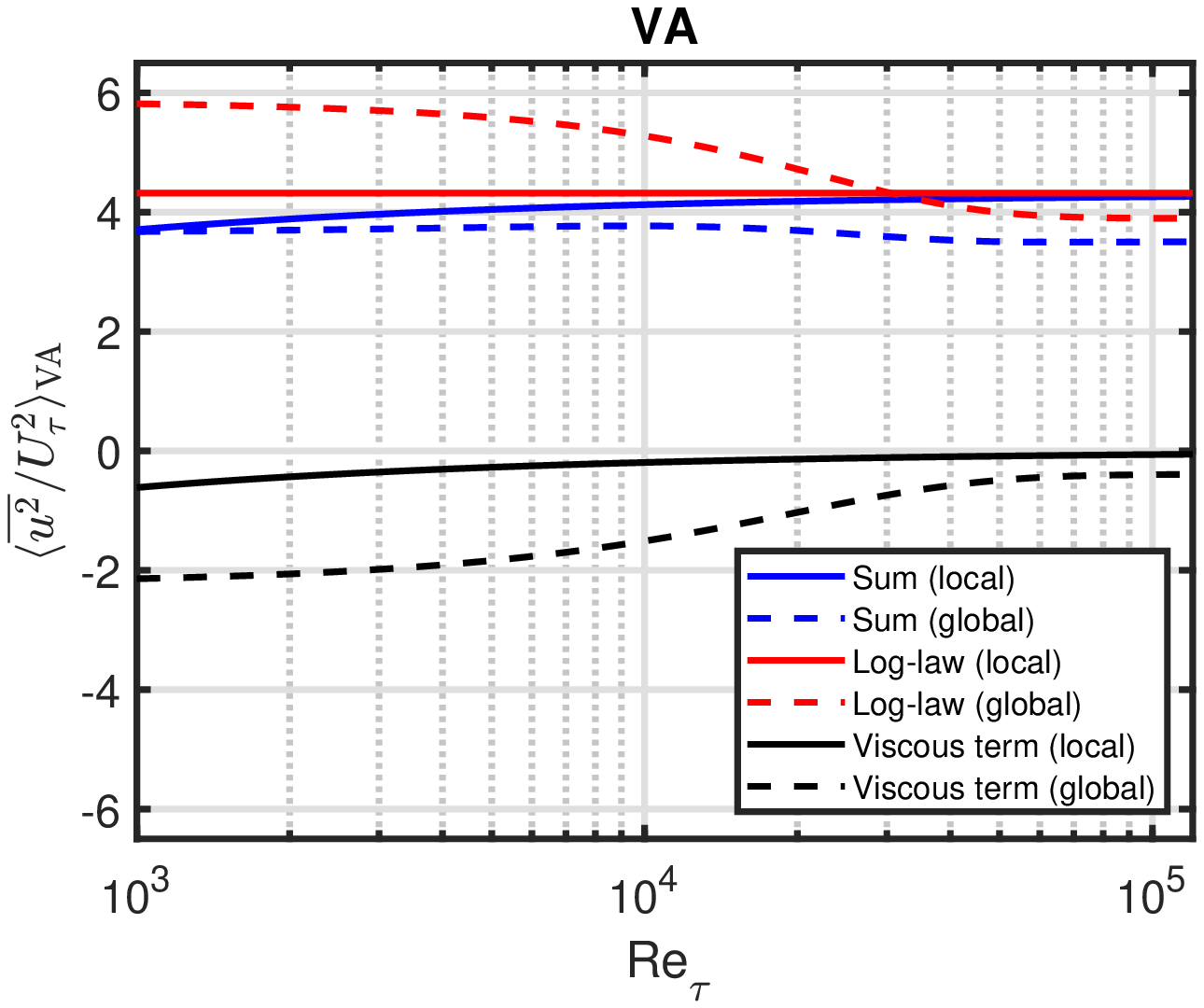}
\caption{Decomposition of the VA average as a function of $Re_{\tau}$. Left-hand plot: Individual points, right-hand plot: Comparison of local and global fits.}
\label{fig:fluc_VA_avrg}
\end{figure}

The decomposition of the fluctuations into a log-law and a viscous term is shown in the left-hand plots of Figures \ref{fig:fluc_AM_avrg} and \ref{fig:fluc_VA_avrg}. A comparison of the corresponding log-law and viscous terms for local and global fits is available in the right-hand plots of Figures \ref{fig:fluc_AM_avrg} and \ref{fig:fluc_VA_avrg}.

\begin{figure}[!ht]
\centering
\includegraphics[width=6.5cm]{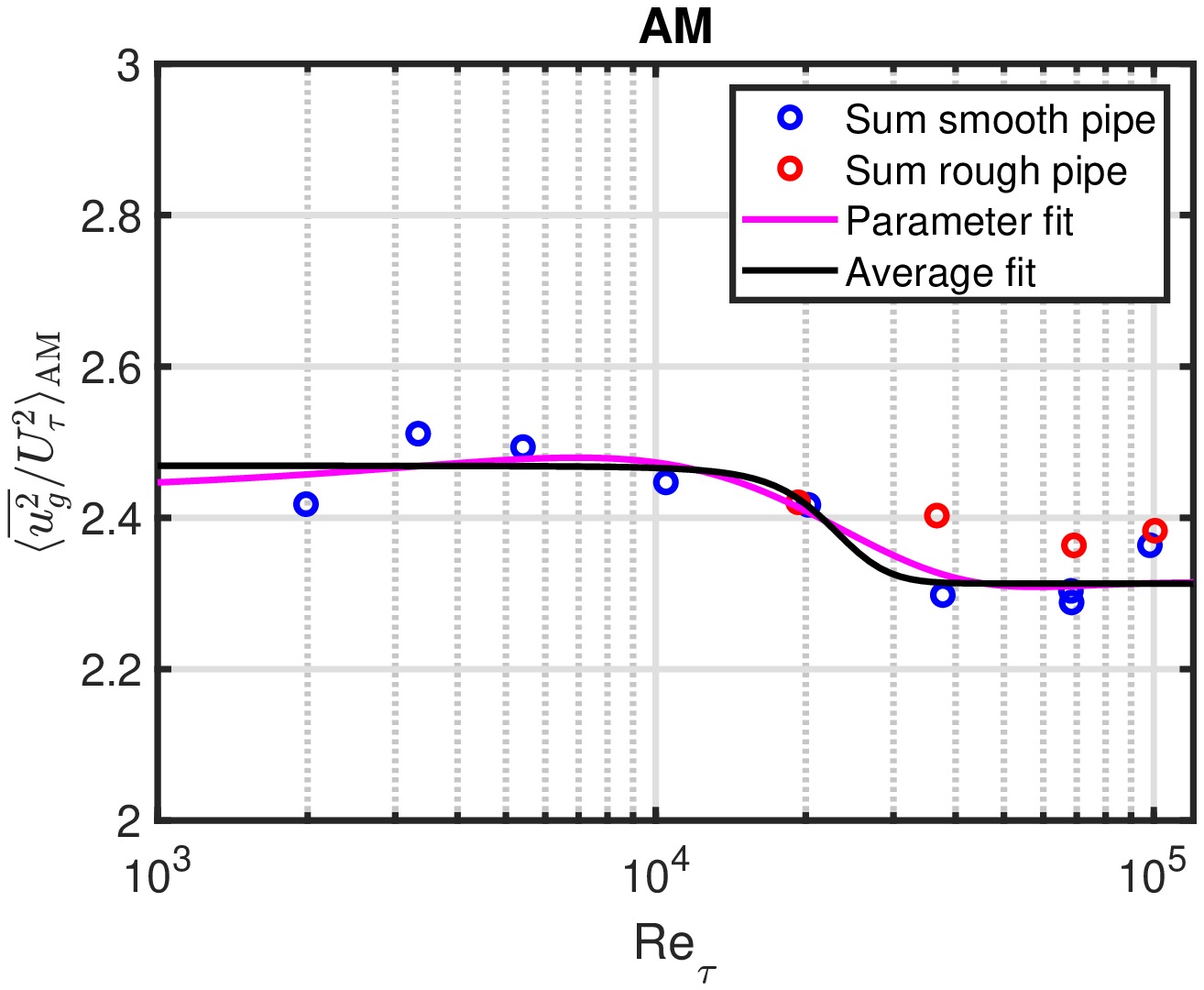}
\hspace{0.3cm}
\includegraphics[width=6.5cm]{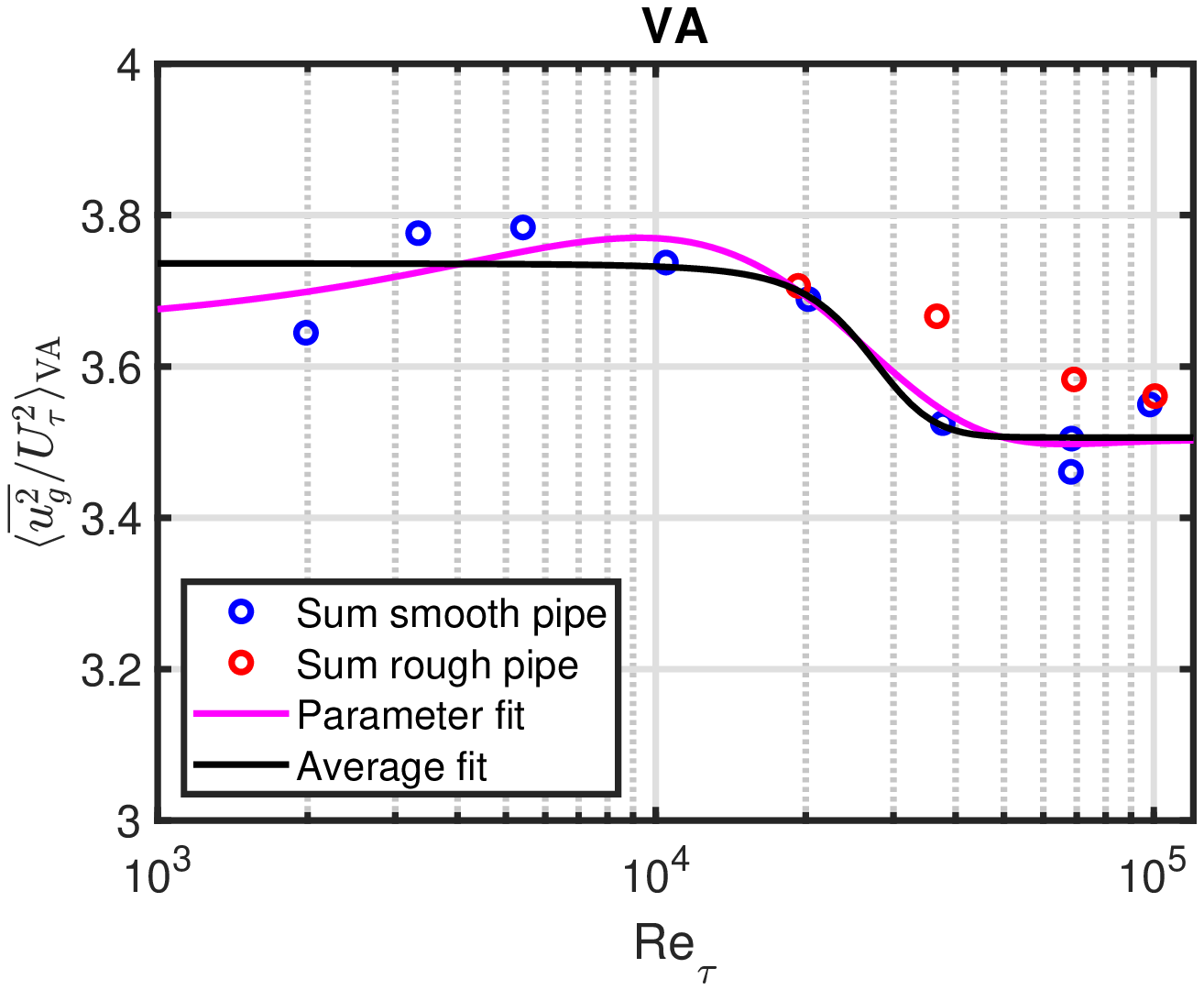}
\caption{AM (left-hand plot) and VA (right-hand plot) average as a function of $Re_{\tau}$. Fit to the sum using Equation (\ref{eq:Q_fit}), both for the parameters and the average.}
\label{fig:AM_local_global}
\end{figure}

Hyperbolic tangent fits, both using the individual parameter fits and a fit to the average, are included in Figure \ref{fig:AM_local_global}.

\clearpage

%\section*{References}
\label{sec:refs}

\end{document}